\documentclass[11pt]{article}
\baselineskip
\pdfoutput=1
\usepackage{amsmath}
\usepackage{amssymb}
\usepackage{graphicx}
\usepackage[T1]{fontenc}
\usepackage{cite}
\usepackage{collref}
\usepackage{tabls}
\usepackage{hyperref}
\newcommand{\SU}{\mathrm{SU}}
\newcommand{\dd}{{\rm{d}}}
\newcommand{\real}{{\rm Re\,}}
\newcommand{\Tr}{{\rm Tr\,}}
\newcommand{\kB}{k_{\mbox{\tiny{B}}}}
\newcommand{\Zin}{Z_{\mbox{\tiny{in}}}}
\newcommand{\Zfin}{Z_{\mbox{\tiny{fin}}}}
\newcommand{\phiin}{\phi_{\mbox{\tiny{in}}}}
\newcommand{\tin}{t_{\mbox{\tiny{in}}}}
\newcommand{\tfin}{t_{\mbox{\tiny{fin}}}}
\newcommand{\Tin}{T_{\mbox{\tiny{in}}}}
\newcommand{\lambdain}{\lambda_{\mbox{\tiny{in}}}}
\newcommand{\lambdafin}{\lambda_{\mbox{\tiny{fin}}}}
\newcommand{\betain}{\beta_{\mbox{\tiny{in}}}}
\newcommand{\betafin}{\beta_{\mbox{\tiny{fin}}}}
\newcommand{\ntraj}{n_{\mbox{\tiny{traj}}}}
\newcommand{\nint}{n_{\mbox{\tiny{int}}}}
\newcommand{\ntherm}{n_{\mbox{\tiny{therm}}}}
\newcommand{\Tc}{T_{\mbox{\tiny{c}}}}
\newcommand{\alphastrong}{\alpha_{\mbox{\tiny{s}}}}
\newcommand{\nconf}{n_{\mbox{\tiny{conf}}}}

\begin{document}

\begin{titlepage}
\begin{flushright} 
DESY-18-001
\end{flushright}
\vskip0.5cm
\renewcommand\thefootnote{\mbox{$\fnsymbol{footnote}$}}
\begin{center}
{\Large\bf QCD thermodynamics from lattice calculations with non-equilibrium methods: The $\SU(3)$ equation of state}
\end{center}
\vskip1.3cm
\centerline{Michele~Caselle$^{a,b,c}$, Alessandro~Nada$^{a,c,d}$, and Marco~Panero$^{a,c}$}
\vskip1.5cm
\centerline{\sl $^a$Department of Physics and $^b$Arnold-Regge Center, University of Turin, and $^c$INFN, Turin}
\centerline{\sl Via Pietro Giuria 1, I-10125 Turin, Italy}
\vskip0.5cm
\centerline{\sl $^d$NIC, DESY}
\centerline{\sl Platanenallee 6, D-15738 Zeuthen, Germany}
\vskip0.5cm
\begin{center}
{\sl  E-mail:} \hskip 1mm \href{mailto:caselle@to.infn.it}{{\tt caselle@to.infn.it}}, \href{mailto:alessandro.nada@desy.de}{{\tt alessandro.nada@desy.de}},\\ \href{mailto:marco.panero@unito.it}{{\tt marco.panero@unito.it}}
\end{center}
\vskip1.0cm
\begin{abstract}
A precise lattice determination of the equation of state in $\SU(3)$ Yang-Mills theory is carried out by means of a simulation algorithm, based on Jarzynski's theorem, that allows one to compute physical quantities in thermodynamic equilibrium, by driving the field configurations of the system out of equilibrium. The physical results and the computational efficiency of the algorithm are compared with other state-of-the-art lattice calculations, and the extension to full QCD with dynamical fermions and to other observables is discussed.
\end{abstract}

\end{titlepage}

\section{Introduction and motivation}
\label{sec:intro}

The phenomenology of the strong interaction at high temperatures and/or densities remains one of the most interesting (yet somehow elusive) research areas in the physics of elementary particles. As nicely summarized by B.~M\"uller in his lecture at the 2013 Nobel Symposium on LHC Physics~\cite{Muller:2013dea}, the novel state of matter produced in nuclear collisions at LHC and RHIC reveals unique features: it is strongly coupled, but highly relativistic; at high temperature it displays the distinctive collective phenomena of a liquid, whereas at low temperatures it turns into a gas of weakly interacting hadrons; while its shear viscosity $\eta$ is nearly $18$ orders of magnitude larger than the one measured for superfluid helium and even $26$ orders of magnitude larger than the one of ultracold atoms~\cite{Schafer:2009dj}, the ratio of the shear viscosity over the entropy density $s$ is actually \emph{lower} than for those substances, and close to the fundamental quantum-mechanical bound $1/(4\pi)$~\cite{Policastro:2001yc}; moreover, it thermalizes in a very short time, close to the limits imposed by causality. Finally, the quark-gluon plasma (QGP) is not simply a ``rearrangement'' of ordinary nuclear matter: rather, it ``creates'' its own ground state, in which two characterizing features of the hadronic world, color confinement and dynamical chiral-symmetry breaking, are lost.

At the temperatures reached in present heavy-ion-collision experiments---which, when expressed in natural units $\hbar=c=\kB=1$, are of the order of hundreds of MeV~\cite{Braun-Munzinger:2015hba}---the QGP is strongly coupled: this demands a theoretical investigation by non-perturbative tools, and the regularization of quantum chromodynamics (QCD) on a Euclidean lattice~\cite{Wilson:1974sk} is the tool of choice for this purpose. Over the past few years, several physical observables relevant for finite-temperature QCD have been studied on the lattice (see refs.~\cite{Philipsen:2012nu, Ding:2015ona, Meyer:2015wax} for reviews): one of the most prominent among them is the QCD equation of state~\cite{Borsanyi:2013bia, Bazavov:2014pvz}, which determines the evolution of the Universe shortly after the Big~Bang, as well as the evolution of the matter produced in the ``little bang'' at ultrarelativistic nuclear colliders.

While state-of-the-art results for the QCD equation of state, obtained by different collaborations using slightly different types of lattice discretizations, are now consistent with each other, it is worth remarking that such computations still require large computational power, and the multiple extrapolations to the physical limit are far from trivial. For example, in the standard ``integral'' method~\cite{Engels:1990vr}, the fact that quantum fluctuations at the lattice cutoff scale induce a strong ultraviolet divergence in the free energy associated with the QCD partition function, implies that bulk quantities at thermal equilibrium, such as the pressure $p$ at a finite temperature $T$, have to be extracted by subtracting the corresponding quantities evaluated in vacuum, and are encoded in numbers that scale like $O(a^D)$ ($a$ being the lattice spacing, and $D$ the Euclidean spacetime dimension, i.e. four): this constrains the values of $a$ that can be probed in these simulations and, as a consequence, the control over systematic uncertainties affecting the extrapolation to the continuum. Similarly, in simulations with staggered fermions, residual taste-symmetry-breaking effects can have an impact on the extrapolation of the quark masses to the physical limit.

Due to these challenges, in the past few years there has been renovated interest in alternative methods to compute the equation of state. In particular, we would like to mention two recent studies, based upon the gradient flow~\cite{Kitazawa:2016dsl} (see also ref.~\cite{Asakawa:2013laa}, which reported the first calculation of thermodynamic quantities using this method, and the very recent ref.~\cite{Hirakida:2018uoy}, for an application in $\SU(2)$ Yang-Mills theory) and on the formulation of the theory in a moving reference frame~\cite{Giusti:2016iqr}: both of them have been successfully tested in $\SU(3)$ Yang-Mills theory without quarks, and can be extended to full QCD without major obstructions~\cite{Kanaya:2016rkt, DallaBrida:2017sxr}. The thermal properties of a purely gluonic theory, albeit not relevant for a quantitative comparison with experiments, can reveal important universal features, shared by theories with different gauge symmetry~\cite{Boyd:1996bx, Lucini:2005vg, Lucini:2012wq, Panero:2008mg, Borsanyi:2012ve, Francis:2015lha, Kitazawa:2016dsl, Asakawa:2013laa, Giusti:2016iqr, Engels:1988ph, Engels:1994xj, Lucini:2002ku, Lucini:2003zr, Bursa:2005yv, Bringoltz:2005rr, Bringoltz:2005xx, Pepe:2006er, Cossu:2007dk, Umeda:2008bd, Meyer:2009tq, Panero:2009tv, Datta:2009jn, Wellegehausen:2009rq, Datta:2010sq, Mykkanen:2012ri, Lucini:2012gg, Giusti:2014ila, Bruno:2014rxa, Bonati:2015uga, Hajizadeh:2017jsw, Giudice:2017dor} and/or in different dimensions~\cite{Christensen:1991rx, Holland:2005nd, Holland:2007ar, Liddle:2008kk, Bialas:2008rk, Caselle:2011fy, Caselle:2011mn, Bialas:2012qz}, and, by virtue of the limited computational power required for their numerical Monte~Carlo simulation, provide a useful benchmark for new algorithms.

In this manuscript, we present yet another method to compute the $\SU(3)$ equation of state, which is based on Jarzynski's theorem~\cite{Jarzynski:1996oqb, Jarzynski:1997ef}: as will be discussed in detail in section~\ref{sec:Jarzynski_equality}, this theorem encodes an exact relation between the ratio of the partition functions associated with two different ensembles (which, in this case, are defined as those of the theory at two different temperatures) to an exponential average of the work done on the system during \emph{a non-equilibrium transformation} driving it from one ensemble to the other. As will be discussed in detail below, calculations of the pressure based on this technique still require the subtraction of ultraviolet vacuum contributions, as with the integral method; however, they strongly reduce the computational costs associated with thermalization, since, in contrast to the integral method, only the field configurations at the first temperature in each trajectory need to be thermalized. Jarzynski's theorem is closely related to a set of powerful mathematical identities in non-equilibrium statistical physics, which have been developed since the 1990's~\cite{Evans:1993po, Evans:1993em, Gallavotti:1994de, Gallavotti:1995de, Crooks:1997ne, Crooks:1999ep, Crooks:1999pe, Ritort:2004wf, MariniBettoloMarconi:2008fd}. A first example of application of Jarzynski's theorem in numerical simulations of lattice gauge theory was presented in ref.~\cite{Caselle:2016wsw}, but the technique is quite general and versatile, and can be used for a variety of different lattice QCD problems (at zero or at finite temperature). In section~\ref{sec:results}, after laying out the setup of our numerical calculations, we report a set of high-precision results for the $\SU(3)$ equation of state obtained using this method, along with a detailed discussion of the underlying physics, and with a comparison to studies based on different methods~\cite{Borsanyi:2012ve, Giusti:2016iqr, Kitazawa:2016dsl}. Section~\ref{sec:discussion} is devoted to a discussion of the computational efficiency of our method and to some concluding remarks. A summary of this work has been reported in ref.~\cite{Nada:2017xmz}.

\section{Jarzynski's equality}
\label{sec:Jarzynski_equality}

In this section, after stating Jarzynski's theorem, we first demonstrate it in a Hamiltonian-evolution framework, following ref.~\cite{Jarzynski:1996oqb}, in subsection~\ref{subsec:Hamiltonian-evolution_proof}. Then, in subsection~\ref{subsec:master-equation_proof}, we present a different derivation~\cite{Jarzynski:1997ef}, based on a master-equation formalism, which is more directly relevant for a practical implementation in Monte~Carlo calculations.

Jarzynski's equality~\cite{Jarzynski:1996oqb, Jarzynski:1997ef} is a theorem in statistical mechanics, that relates equilibrium and non-equilibrium quantities.

Consider a classical statistical system, which depends on a set of parameters $\lambda$ (defined in a space $\Lambda$), and let $H$ denote its Hamiltonian, which is a function of the degrees of freedom $\phi$. When the system is in thermal equilibrium at temperature $T$, the partition function, defined as
\begin{equation}
\label{partition_function}
Z = \sum_{\left\{ \phi \right\}} \exp \left( - \frac{H}{T} \right)
\end{equation}
(where $\sum_{\left\{ \phi \right\}}$ denotes the sum over all possible $\phi$ configurations, and, depending on the nature of $\phi$ and on the theory, may be a finite or an infinite sum, a multiple integral, or a suitably defined functional integral), is related to the Helmholtz free energy $F$ via $Z=\exp(-F/T)$. In eq.~(\ref{partition_function}), both the partition function and the free energy, like $H$, are functions of $\lambda$. Let $\lambdain$ and $\lambdafin$ denote two distinct values of $\lambda$ in parameter space, and let $Z_{\lambdain}$ and $Z_{\lambdafin}$ denote the partition functions of the system in thermodynamic equilibrium, when its parameters take values $\lambda=\lambdain$ and $\lambda=\lambdafin$, respectively. For a given physical observable $\mathcal{O}$, let $\langle \mathcal{O} \rangle_\lambda$ denote the statistical average of $\mathcal{O}$ in thermal equilibrium in the ensemble with parameters fixed to $\lambda$.

Consider now the situation in which the parameters of the system are varied as a function of time $t$ in a certain interval (which can be either finite or infinite) of extrema $\tin$ and $\tfin$, according to some, arbitrary but well-specified, function $\lambda(t)$ (or ``protocol'' for the parameter evolution), with $\lambda(\tin)=\lambdain$ and $\lambda(\tfin)=\lambdafin$. Assume that, starting from an initial equilibrium configuration at $t=\tin$, the parameters are let evolve in time, according to the $\lambda(t)$ function; accordingly, the dynamical variables $\phi$ respond to the variation in the $\lambda$ parameters, and themselves evolve in time, spanning a trajectory in the field-configuration space. In general, the configurations at all $t>\tin$ are not thermalized, i.e. the $\lambda(t)$ parameter evolution drives the system out of equilibrium (except when $\tfin-\tin$ is infinite, so that the switching process is infinitely slow). Let $W$ denote the total work done on the system during its evolution from $\tin$ to $\tfin$; since the system is driven out of equilibrium, the mean value of the work $\overline{W}$ obtained by averaging over an ensemble of such transformations, is in general larger than or equal to the free-energy difference $\Delta F = F_{\lambdafin} - F_{\lambdain}$ of equilibrium ensembles with parameters $\lambda=\lambdain$ and $\lambda=\lambdafin$:
\begin{equation}
\label{average_work_Delta_F}
\overline{W} \ge \Delta F.
\end{equation}
Note that $\overline{W}-\Delta F$ is the amount of work dissipated during the parameter switch, which is directly related to the entropy variation, hence the inequality~(\ref{average_work_Delta_F}) is nothing but an expression of the second law of thermodynamics. Also, when the parameter switch is infinitely slow (i.e. for $\Delta t=\tfin-\tin \to \infty$) the system remains in thermodynamic equilibrium throughout the switching process, the transformation is reversible, and the equality sign holds.

However, if one considers the \emph{exponential average} of the work, then it is possible to prove that it is directly related to $\Delta F$ through the following \emph{equality}:
\begin{equation}
\label{Jarzynski_theorem}
\overline{\exp\left(-W/T\right)} = \exp\left(-\Delta F/T\right).
\end{equation}
Eq.~(\ref{Jarzynski_theorem}) is the main statement of Jarzynski's theorem~\cite{Jarzynski:1996oqb}.

Before discussing the proof of eq.~(\ref{Jarzynski_theorem}) for generic $\Delta t$, we observe that when $\Delta t \to \infty$, the equality holds: in this limit, the parameter switch from $\lambdain$ to $\lambdafin$ is infinitely slow, the transformation becomes quasi-static, the system remains in equilibrium for the whole duration of the process, so that the work done on the system is equal to
\begin{equation}
\label{W_in_quasistatic_limit}
W = \int_{\lambdain}^{\lambdafin} \left\langle \frac{\partial H}{\partial \lambda} \right\rangle_{\lambda} \dd \lambda
\end{equation}
for every trajectory interpolating between the initial and final ensembles. Hence, in this limit one has $\overline{W}=W$. Moreover, in this limit one also has $\overline{W}=\Delta F$, thus the left-hand side of eq.~(\ref{Jarzynski_theorem}) can be written as
\begin{equation}
\overline{\exp\left(-W/T\right)} = \exp\left(-\overline{W}/T\right) = \exp\left(-\Delta F/T\right),
\end{equation}
and eq.~(\ref{Jarzynski_theorem}) is trivially recovered.

\subsection{Derivation in a Hamiltonian-evolution framework}
\label{subsec:Hamiltonian-evolution_proof}

To prove eq.~(\ref{Jarzynski_theorem}) for finite $\Delta t$, let us first consider the case in which the system is initially in thermal equilibrium with a heat reservoir at temperature $T$, but is isolated from it during the switching process from $\tin$ to $\tfin$. Then, one can express the average over the ensemble of trajectories appearing on the left-hand side of eq.~(\ref{Jarzynski_theorem}) in terms of the time-dependent probability density in the space of configurations, that we denote as $\rho=\rho(\phi,t)$. Given that at $t=\tin$ the system is in thermal equilibrium at temperature $T$, $\rho$ satisfies the initial condition $\rho(\phi,\tin)=\exp\left[ - H_{\lambda(\tin)}(\phi)/T\right]/Z_{\lambda(\tin)}$; moreover, since the system is in isolation during the switching process, the time evolution of $\rho$ at $t>\tin$ is given by Liouville's equation $\dot{\rho}=\left\{ H_{\lambda} , \rho \right\}$, where the quantity appearing on the right-hand side is the Poisson bracket of $H_{\lambda}$ and $\rho$. The evolution law expressed by Liouville's equation is fully deterministic, and a one-to-one mapping exists between each configuration at a generic time $t$ and a configuration $\phiin$ at the initial time $t=\tin$. As a consequence, the work accumulated along a trajectory going through a configuration $\phi$ at a generic time $t$ is well-defined and equal to 
\begin{equation}
\label{accumulated_work}
w(\phi,t) = H_{\lambda(t)}(\phi) - H_{\lambda(\tin)}(\phiin)= \int_{\tin}^t \frac{\partial H_{\lambda}}{\partial \lambda} \dot{\lambda} \dd \tau.
\end{equation}
Thus, the work accumulated during the evolution starting from $t=\tin$ and leading to a final configuration $\phi$ at $t=\tfin$ is simply $w(\phi,\tfin)$, and the average appearing on the left-hand side of eq.~(\ref{Jarzynski_theorem}) can be expressed as
\begin{equation}
\label{average_expression_I}
\overline{\exp\left(-W/T\right)} = \sum_{\left\{ \phi \right\}} \rho(\phi,\tfin) \exp\left[-w(\phi,\tfin)/T\right].
\end{equation}
Liouville's theorem implies the conservation of the trajectory density in phase space: hence, $\rho(\phi,\tfin)=\rho(\phiin,\tin)=\exp\left[ - H_{\lambda(\tin)}(\phiin)/T\right]/Z_{\lambda(\tin)}$, so that eq.~(\ref{average_expression_I}) can be rewritten as
\begin{eqnarray}
\label{average_expression_II}
\overline{\exp\left(-W/T\right)} &=& \frac{1}{Z_{\lambda(\tin)}} \sum_{\left\{ \phi \right\}} \exp\left[ - \frac{H_{\lambda(\tin)}(\phiin)}{T}\right] \exp\left[-\frac{H_{\lambda(t)}(\phi) - H_{\lambda(\tin)}(\phiin)}{T}\right] \nonumber \\
&=& \frac{1}{Z_{\lambda(\tin)}} \sum_{\left\{ \phi \right\}} \exp\left[ - \frac{H_{\lambda(t)}(\phi)}{T}\right] = \frac{Z_{\lambda(\tfin)}}{Z_{\lambda(\tin)}} = \exp\left(-\Delta F/T\right).
\end{eqnarray}

If the system remains coupled to a heat reservoir doing the parameter switch (and the coupling of the system to the reservoir is sufficiently small), then this argument can be repeated for the union of the system and the reservoir, which can be thought of as a larger system, that remains isolated during the process. Then, the work performed on the system equals the difference of the total energy, evaluated on the final and on the initial configuration. This difference does not depend on the switching time, therefore it can be evaluated in the $\Delta t \to \infty$ limit, in which, as we discussed above, eq.~(\ref{Jarzynski_theorem}) holds. Actually, one can prove that the assumption of weak coupling between the system and the reservoir can be relaxed, if the reservoir is mimicked by a Nos\'e-Hoover thermostat~\cite{Nose:1984au, Hoover:1985zz} or a Metropolis algorithm, as is the case in Monte~Carlo simulations.

\subsection{Derivation in the master-equation formalism}
\label{subsec:master-equation_proof}

Eq.~(\ref{Jarzynski_theorem}) can also be derived using a master-equation approach, and assuming a completely stochastic (rather than deterministic) evolution for the trajectory~\cite{Jarzynski:1997ef}. Here and in the following, we will use the symbol $\phi$ to denote a field configuration of the system, and $\phi(t)$ will denote a field configuration at time $t$. Here, the time evolution of $\phi$ is assumed to be given by a stochastic process; as a result of this stochastic process, the field configuration changes with time, and, following ref.~\cite{Jarzynski:1997ef}, we will call this process a ``trajectory'' in the space of the possible configurations of the system. Let $P(\phi^\prime,t|\phi,t+\Delta t)$ denote the conditional probability of finding a field configuration $\phi$ at time $t+\Delta t$, given that the system was in configuration $\phi^\prime$ at time $t$, and define the instantaneous transition rate from $\phi^\prime$ to $\phi$ as
\begin{equation}
\label{R_definition}
R_\lambda(\phi^\prime,\phi) = \lim_{\Delta t \to 0^+} \frac{\partial}{\partial (\Delta t)} P(\phi^\prime,t|\phi,t+\Delta t).
\end{equation}
Note that this quantity depends on time only through the time-dependence of $\lambda$. Consider now an ensemble of stochastic, Markovian temporal evolutions (or trajectories) of the system, given a certain, fixed time-evolution of its parameters, $\lambda(t)$: the distribution density of these trajectories in the space of configurations of the system, denoted as $f(\phi,t)$, obeys
\begin{equation}
\label{master_equation}
\frac{\partial}{\partial t} f(\phi,t) = \sum_{\left\{ \phi^\prime \right\}} f(\phi^\prime,t) R_\lambda(\phi^\prime,\phi)=\hat{R}_\lambda f,
\end{equation}
where the last equality is the \emph{definition} of the $\hat{R}_\lambda$ operator. If $\lambda$ does not depend on time, then the formal solution of eq.~(\ref{master_equation}), with the boundary condition that at $t=\tin$ the distribution density equals $f_{\mbox{\tiny{in}}}(\phi)$, can be written as
\begin{equation}
\label{master_equation_solution}
f(\phi,t) = \exp\left[ (t-\tin) \hat{R}_\lambda \right] f_{\mbox{\tiny{in}}}(\phi).
\end{equation}
In this case, $\phi(t)$ reduces to a standard, stationary Markov process: then, the distribution density $f(\phi,t)$ becomes time-independent and the left-hand side of eq.~(\ref{master_equation}) vanishes. Thus, the Markov process generates an ensemble of configurations distributed according to the canonical Boltzmann distribution for a system with Hamiltonian $H_\lambda$ at temperature $T$, i.e. $f(\phi,t) \propto \exp[-H_\lambda(\phi)/T]$, and
\begin{equation}
\label{detailed_balance}
\sum_{\left\{ \phi^\prime \right\}} \exp[-H_\lambda(\phi^\prime)/T] R_\lambda(\phi^\prime,\phi)=0.
\end{equation}
Eq.~(\ref{detailed_balance}) means that the canonical distribution is preserved by the Markov process under consideration. Note that, if the Markov process satisfies detailed balance, i.e. if
\begin{equation}
\label{actual_detailed_balance}
\frac{\exp[-H_\lambda(\phi^\prime)/T]}{\exp[-H_\lambda(\phi)/T]} = \frac{R_\lambda(\phi,\phi^\prime)}{R_\lambda(\phi^\prime,\phi)},
\end{equation}
then eq.~(\ref{detailed_balance}) follows: this can be easily proven by expressing $\exp[-H_\lambda(\phi^\prime)/T] R_\lambda(\phi^\prime,\phi)$ in terms of $\exp[-H_\lambda(\phi)/T]$ and $R_\lambda(\phi,\phi^\prime)$ using eq.~(\ref{actual_detailed_balance}), and then summing (or integrating) over the $\phi^\prime$ values. The converse is in general not true, but, given that the distinction between eq.~(\ref{detailed_balance}) and eq.~(\ref{actual_detailed_balance}) is immaterial for our present discussion, for the sake of simplicity we will nevertheless refer to eq.~(\ref{detailed_balance}) as to the ``detailed-balance condition'', as was done in ref.~\cite{Jarzynski:1997ef}.\footnote{One can also assume the stronger condition that, when $t - \tin \to \infty$, the Markov process always generates a canonical Boltzmann distribution, i.e. that for any, arbitrary, initial distribution $f_{\mbox{\tiny{in}}}(\phi)$:
\begin{equation}
\label{thermalization}
\lim_{(t - \tin) \to \infty} \exp\left[ (t-\tin) \hat{R}_\lambda \right] f_{\mbox{\tiny{in}}}(\phi)=\frac{\exp[-H_\lambda(\phi)/T]}{\sum_{\left\{ \phi \right\}} \exp[-H_\lambda(\phi)/T]},
\end{equation}
so that, for sufficiently long times, the Markov process always leads to \emph{thermalization} of any distribution. Note that eq.~(\ref{thermalization}) is stronger than and implies eq.~(\ref{detailed_balance}). For our present purposes, however, only eq.~(\ref{detailed_balance}) is needed.}

Let us assume that the initial distribution at time $t=\tin$ is a canonical one, $f_{\mbox{\tiny{in}}}(\phi) \propto \exp[-H_{\lambda(\tin)}(\phi)/T]$, let $Q(\phi,t)$ denote the average value of $\exp[-w(\phi,t)/T]$ over all trajectories going through a particular configuration $\phi$ at a generic time $t$. Introducing the distribution defined as
\begin{equation}
\label{g_definition}
g(\phi,t)=f(\phi,t)Q(\phi,t),
\end{equation}
the average of $\exp(-W/T)$ over all trajectories can be expressed as
\begin{equation}
\label{average_exponential_work_I}
\overline{\exp\left(-W/T\right)} = \sum_{\left\{ \phi \right\}} g(\phi,\tfin).
\end{equation}
From its definition by eq.~(\ref{g_definition}), it is easy to see that the time derivative of $g$ is given by
\begin{equation}
\label{g_equation}
\frac{\partial g}{\partial t} = \frac{\partial f}{\partial t}Q + f \frac{\partial Q}{\partial t} = \hat{R}_\lambda f Q - f \frac{\partial H_{\lambda}}{\partial \lambda} \frac{\dot{\lambda}}{T} Q = \left( \hat{R}_\lambda - \frac{\partial H_{\lambda}}{\partial \lambda} \frac{\dot{\lambda}}{T} \right) g.
\end{equation}
In particular, the third equality appearing in eq.~(\ref{g_equation}) can be proven by imagining that $\phi(t)$ represents the ``motion of a particle with a time-dependent mass $\exp[-w(\phi,t)/T]$'' (this motion is supposed to take place in the space of configurations), so that $Q(\phi,t)$ can then be interpreted as the ``average mass'' of the particles that at time $t$ go through $\phi(t)$, and $g(\phi,t)$ represents the ``average mass density'' of the particles that go through $\phi$ at time $t$. The time dependence of such ``average mass density'' would then be induced by two terms: first, the one due to the the ``flow'' of these ``particles'', which is encoded in eq.~(10), and, second, by the fact that the particle ``mass'' $m(t)=\exp[-w(\phi,t)/T]$ varies with time, and $\dot{m}(t)=-[w(\phi,t)/T]m(t)$. The time derivative of $g$ is then given by the sum of these two terms, which yields eq.~(\ref{g_equation}). For another derivation of eq.~(\ref{g_equation}), see ref.~\cite[appendix~A]{Jarzynski:1997ef}. Note that eqs.~(\ref{accumulated_work}) and~(\ref{g_definition}) imply that, at $t=\tin$:
\begin{equation}
\label{g_boundary_condition}
g(\phi,\tin)=f(\phi,\tin)=\frac{\exp[-H_{\lambda(\tin)}(\phi)/T]}{\sum_{\left\{ \phi \right\}} \exp[-H_{\lambda(\tin)}(\phi)/T]}=\frac{\exp[-H_{\lambda(\tin)}(\phi)/T]}{\Zin},
\end{equation}
where we used the fact that the initial distribution is a canonical one.

According to eq.~(\ref{detailed_balance}), $\hat{R}_\lambda$ annihilates $\mathcal{N}\exp(-H_\lambda/T)$ (where $\mathcal{N}$ is an arbitrary constant factor), hence:
\begin{equation}
\left( \frac{\partial }{\partial t} -\hat{R}_\lambda + \frac{\partial H_{\lambda}}{\partial \lambda} \frac{\dot{\lambda}}{T} \right) \mathcal{N}\exp(-H_\lambda/T) = 0,
\end{equation}
which means that $\mathcal{N}\exp(-H_\lambda/T)$ is solution to eq.~(\ref{g_equation}). The solution consistent with the boundary condition specified by eq.~(\ref{g_boundary_condition}) has $\mathcal{N}=1/\Zin$, so that
\begin{equation}
\label{g_solution}
g(\phi,t)= \frac{\exp[-H_\lambda(t)(\phi)/T]}{\Zin}.
\end{equation}
Plugging eq.~(\ref{g_solution}), evaluated at $t=\tfin$, into eq.~(\ref{average_exponential_work_I}), one finally obtains
\begin{equation}
\label{average_exponential_work_II}
\overline{\exp\left(-W/T\right)} = \frac{1}{\Zin} \sum_{\left\{ \phi \right\}} \exp[-H_{\lambda(\tfin)}(\phi)/T] = \frac{\Zfin}{\Zin},
\end{equation}
which proves Jarzynski's theorem.

Note that, even though the distribution of $\phi$ is a canonical one only at $t=\tin$, in the last term of eq.~(\ref{average_exponential_work_II}) the canonical partition function of the system at the final value of $\lambda$ appears, and that this equation relates a genuinely out-of-equilibrium quantity (the average appearing in the first term) to a ratio of equilibrium quantities.

This proof of Jarzynski's equality provides a natural way to implement a numerical evaluation of the free-energy difference appearing on the right-hand side of eq.~(\ref{Jarzynski_theorem}) by Monte~Carlo simulation:\footnote{A related idea underlies the annealed-importance-sampling technique~\cite{Neal:2001ai}: we thank Martin~Hasenbusch for discussions on this issue.} having defined a parameter evolution $\lambda(t)$, with $\tin \le t \le \tfin$, that interpolates between the initial and final ensembles, and starting from a canonical distribution of configurations, one can drive the system out of equilibrium by varying $\lambda$ as a function of Monte~Carlo time, letting the configurations evolve according to any Markov process that satisfies the detailed-balance condition expressed by eq.~(\ref{detailed_balance}), and compute $\exp(-W/T)$ during this process. The average expressed by the bar notation on the left-hand side of eq.  ~(\ref{Jarzynski_theorem}) is then obtained by averaging over a sufficiently large number of such trajectories. This is the numerical strategy that we use in this work, in which the Euclidean action $S$ plays the r\^ole of $H/T$.

We close this section with a word of caution. The computational efficiency of this method may strongly depend on the properties of the system under consideration: in particular, physical systems with a very large number of degrees of freedom (such as quantum field theories regularized on a spacetime lattice) have sharply peaked statistical distributions, hindering an accurate sampling of the configuration-space regions that contribute mostly to $\overline{\exp(-W/T)}$. If the different values of $W$ in different trajectories are much larger than the scale of typical thermal fluctuations (or of typical quantum fluctuations, for lattice simulations of quantum field theory), then $\overline{\exp(-W/T)}$ is dominated by configurations in which the value of $W$ is much smaller than $\overline{W}$, and an accurate determination of $\overline{\exp(-W/T)}$ may require a prohibitively large number of trajectories. Note, however, that, in the numerical calculation of free-energy differences by eq.~(\ref{Jarzynski_theorem}), there exists a remarkable difference in the r\^oles of the initial and final ensembles: one assumes that the initial configurations are thermalized, while the field values at all $t > \tin$ (including, in particular, at $t=\tfin$) are out of equilibrium. This asymmetry between the initial and target ensembles implies that, if the Monte~Carlo determination of $\Delta F$ is biased by effects due to limited statistics, then carrying out the same calculation in the opposite direction will, in general, give a result different from $-\Delta F$. Conversely, verifying that a ``direct'' and a ``reverse'' computation give consistent results, provides a powerful test of the correctness of the calculation. This is a test that all results of our present work pass with success.

\section{Lattice calculation of the $\SU(3)$ equation of state}
\label{sec:results}

In this work, we investigate the behavior of QCD at finite temperature, and compute the equation of state via lattice simulations using an algorithm based on Jarzynski's equality eq.~(\ref{Jarzynski_theorem}).

In particular, we focus on the pure-glue sector, which captures the main feature of thermal QCD at the qualitative level: the existence of a confining phase at low temperatures, in which the physical states are massive color singlets, and a deconfined phase at high temperatures, in which chromoelectrically charged, light, elementary particles interact with each other through screened, long-range interactions.\footnote{We also remind the reader of some notable differences between pure-glue $\SU(3)$ Yang-Mills theory and real-world QCD with dynamical quarks. In particular, in the pure-glue theory, the confining and deconfined phases are separated by a first-order phase transition taking place at a critical temperature $\Tc$ which, when converted into physical units, is about $270$~MeV. By contrast, in QCD with physical quarks, the change of state from the confining to the deconfined regimes is rather a smooth crossover, taking place at a lower temperature, around $160$~MeV. However, it has been recently argued that the pure Yang-Mills dynamics could nevertheless be relevant for certain aspects of the physics of heavy-ion collisions' experiments~\cite{Stoecker:2015zea, Stocker:2015nka}.} Thermal screening of both electric and magnetic field components is, indeed, a characterizing feature of the deconfined phase of non-Abelian gauge theories, which defines it as a ``plasma''. Asymptotic freedom implies that, when the temperature $T$ is very high, the physical coupling $g$ at the scale of thermal excitations, $O(T)$, becomes small; in this limit, chromoelectric fields are screened on distances inversely proportional to $gT$, while chromomagnetic fields are screened on lengths inversely proportional to $g^2T$, so that the theory develops a well-defined hierarchy of scales, between ``hard'' (of the order of $T$), ``soft'' (of the order of $gT$), and ``ultra-soft'' (of the order of $g^2T$) modes, and this separation of scales allows for a systematic treatment in terms of effective theories~\cite{Ginsparg:1980ef, Appelquist:1981vg, Nadkarni:1982kb, Braaten:1989mz, Braaten:1995cm, Kajantie:1995dw, Braaten:1995jr, Kajantie:1997tt, Kajantie:2002wa, Kajantie:2003ax}. The appearance of the soft and ultra-soft scales is due to the existence of infra-red divergences, which lead to a breakdown of the correspondence between the number of loops in Feynman diagrams and the order in $\alphastrong$ in perturbative calculations~\cite{Linde:1980ts, Gross:1980br}, and to the intrinsically non-perturbative nature of long-wavelength modes \emph{at all temperatures}. Moreover, for plasma excitations on the energy scale of the deconfinement temperature, the physical coupling is not very small, so that the deconfined state of matter cannot be reliably modeled as a gas of free partons. 

For these reasons, the study of the equation of state of QCD---or of its gluonic sector, that we are focusing on here---close to deconfinement requires non-perturbative techniques. We carry out this study by discretizing the Euclidean action of $\SU(3)$ Yang-Mills theory on a hypercubic lattice $\Lambda$ of spacing $a$, spatial volume $V=L^3=(aN_s)^3$ and extent $aN_t$ along the compactified Euclidean-time direction, using the Wilson action~\cite{Wilson:1974sk}
\begin{equation}
\label{Wilson_action}
 S = \beta \sum_{x \in \Lambda} \sum_{0 \le \mu < \nu \le 3} \left( 1 - \frac{1}{3} \real \Tr U_{\mu\nu} (x) \right),
\end{equation}
where $\beta=6/g_0^2$, with $g_0$ the bare coupling, and 
\begin{equation}
\label{plaquette}
U_{\mu\nu} (x) = U_\mu (x) U_\nu \left(x+a\hat{\mu}\right) U_{\mu}^\dagger \left(x+a\hat{\nu}\right) U_{\nu}^\dagger (x).
\end{equation}
The partition function of the lattice theory is given by
\begin{equation}
\label{lattice_partition_function}
Z = \int \prod_{x \in \Lambda} \prod_{\mu = 0}^{3} \dd U_\mu(x) \exp \left[ -S (U) \right]
\end{equation}
(where $\dd U_\mu(x)$ is the Haar measure for the $\SU(N)$ matrix defined on the oriented link from site $x$ to site $x+a\hat{\mu}$) and expectation values are defined as
\begin{equation}
\label{lattice_expectation value}
\langle \mathcal{O} \rangle = \frac{1}{Z} \int \prod_{x \in \Lambda} \prod_{\mu = 0}^{3} \dd U_\mu(x) \mathcal{O} \exp \left[ -S (U) \right].
\end{equation}
The integrals on the right-hand side of eq.~(\ref{lattice_expectation value}) are estimated numerically, by Monte~Carlo integration, from a sample of field configurations produced in a Markov chain; our update algorithm combines one heat-bath~\cite{Creutz:1980zw, Kennedy:1985nu} and five to ten over-relaxation steps~\cite{Adler:1981sn, Brown:1987rra} on the link variables of the whole lattice: this defines a ``sweep''. The uncertainties in these simulation results are estimated with the jackknife method~\cite{bootstrap_jackknife_book}.

The physical temperature of the system $T=1/(aN_t)$ is varied by varying $a$, which, in turn, can be continuously tuned by varying $\beta$: to this purpose, we set the scale of our lattice simulations by means of the Sommer scale $r_0$~\cite{Sommer:1993ce} as determined in ref.~\cite{Necco:2001xg}. The critical temperature is related to $r_0$ by $\Tc r_0=0.7457(45)$~\cite{Francis:2015lha}.\footnote{Note that, if $r_0$ is assumed to be of the order of $0.5$~fm (a figure consistent with phenomenological potential models for QCD), then the critical deconfinement temperature in $\SU(3)$ Yang-Mills theory is almost twice as large as in QCD. The fact that deconfinement takes place at lower temperatures for theories with a larger number of colored degrees of freedom in the deconfined phase~\cite{Karsch:2000ps, Lucini:2005vg, Lucini:2012wq} is consistent with a qualitative argument, based on the mismatch between the number of degrees of freedom at low and at high temperatures (see also ref.~\cite{Pepe:2005sz}).}

Our lattice determination of the equation of state rests on the following thermodynamic identity, relating the pressure $p$ to the free energy per unit volume $f=F/V$,
\begin{equation}
\label{pressure}
p = -f = \frac{T}{V} \ln{Z},
\end{equation}
which holds in the thermodynamic limit, $V \to \infty$, and receives negligible corrections for the $L$ and $T$ values considered here~\cite{DeTar:1985kx, Elze:1988zs, Gliozzi:2007jh, Meyer:2009kn, Panero:2008mg}. Following the algorithmic strategy discussed in ref.~\cite{Caselle:2016wsw} for a benchmark study in the $\SU(2)$ theory, we study how the dimensionless $p(T)/T^4$ ratio varies as a function of the temperature, starting from an initial temperature $\Tin$:
\begin{equation}
\label{pressure_difference}
\frac{p(T)}{T^4} - \frac{p(\Tin)}{\Tin^4} = \left(\frac{N_t}{N_s}\right)^3 \ln\frac{Z(T)}{Z(\Tin)}.
\end{equation}
In our simulations, we compute $Z(T)/Z(\Tin)$ by means of Jarzynski's equality, using $\beta$ (by tuning which, as stated above, the temperature can be varied continuously) as the $\lambda$ parameter: $\beta$ is let evolve linearly with the Monte~Carlo time $t$ between the initial ($\betain$) and final ($\betafin$) values corresponding to $\Tin$ and $T$, respectively. More precisely, the $\beta$ interval is discretized in $N$ equal intervals of width $\Delta \beta$, so that $\beta_n=\betain+n(\betafin-\betain)/N = \betain + n \Delta \beta$. Finally, one should remember that the $p(T)$ and $p(\Tin)$ terms appearing on the left-hand side of eq.~(\ref{pressure_difference}) also include contributions from quantum (non-thermal) fluctuations, that depend on the lattice cutoff and diverge in the continuum limit. These contributions can be removed from $p$ by evaluating the quantity appearing on the right-hand side of eq.~(\ref{pressure_difference}) on a lattice of large hypervolume $(aN_0)^4$ at $T=0$ at the same $a$. This leads us to define the physical, renormalized pressure as
\begin{equation}
\label{Jarzynski_pressure}
\frac{p(T)}{T^4} = \frac{p(\Tin)}{T_{\mbox{\tiny{in}}}^4} + \left( \frac{N_t}{N_s} \right)^3 \left[ \ln \overline{\exp \left( - \Delta S_{N_s^3 \times N_t} \right)} 
- \gamma \ln \overline{\exp ( - \Delta S_{N_0^4} )} \right],
\end{equation}
where $\Delta S$ is the variation in Euclidean action during a non-equilibrium trajectory in configuration space:
\begin{equation}
\label{action_variation}
\Delta S = \sum\limits_{n=0}^{N-1} \left\{ S[\beta_{n+1},U(t_n)] - S[\beta_{n},U(t_n)] \right\},
\end{equation}
the $N_s^3 \times N_t$ and $N_0^4$ subscripts respectively indicate that this quantity is evaluated on a finite- or on a zero-temperature lattice, $\gamma=N_s^3 \times N_t/N_0^4$, and the bar denotes the average over a sample of $\ntraj$ non-equilibrium trajectories, which start from canonically distributed initial configurations $\left\{ U(t_0)\right\}$. We remark that, in each of these trajectories, only the initial configuration is thermalized; then one starts driving the system out of equilibrium (by varying its parameters, in this case $\beta$) and all subsequent configurations that are produced during the same trajectory are not let thermalize.

Note that the summands on the right-hand side of eq.~(\ref{action_variation}) are given by the action difference induced by a variation of $\beta$ on the same field configuration. In practice, in order to scan a wide temperature range, from the confining to the deconfined phase, it is more convenient to divide the temperature interval in a number (that we denote as $\nint$) of smaller intervals. In particular, we choose these intervals in such a way that they do not stretch across different phases: this allows us to get rid of potential difficulties that might arise in the numerical sampling of configurations, when the algorithm tries to probe the physics at $T>\Tc$, by driving configurations in the $T<\Tc$ phase out of equilibrium, without letting them thermalize.\footnote{A different computational strategy, that would allow the algorithm to avoid the critical point, consists in deforming the action by adding operators that could turn the deconfinement transition into a crossover (e.g. traces of Wilson lines in the Euclidean-time direction), and varying their coefficients to turn them on only near the critical temperature. This numerical strategy, however, is more complex, and we did not explore it in the present work.} Dividing the $\beta$ range of interest in a different number of intervals that do not cross the phase transition should lead to the same physical results, but $\nint$ has some effect on the numerical efficiency of the simulation algorithm. In particular, smaller values of $\nint$ (i.e. broader intervals in $\beta$) typically require larger values of $N$ and more statistics. On the other hand, larger $\nint$ implies a larger overhead for thermalization of the initial configurations at the start of each transformation (in this work we used $5000$ full thermalization sweeps at $T = 0$ and $15000$ at finite temperature).

We run our simulations on lattices with $N_t=6$, $7$, $8$ and $10$ and for $N_s > 12N_t$ (and typically $N_s \simeq 16N_t$), according to the parameters listed in table~\ref{tab:parameters}, where $\ntraj=10$ throughout, and $\nconf$ denotes the total number of configurations used for each combination of parameters, given by the sum of the $N \cdot \ntraj$ products over all $\nint$ intervals. These calculations were carried out on the A1 Intel Broadwell partition of the MARCONI tier-0 supercomputer of the Italian CINECA consortium, a Lenovo system. The total number of core-hours to produce the numerical results presented in this work was approximately $9 \times 10^5$.

\begin{table}[!htb]
\centering
\begin{tabular}{|c|c|c|c|c|c|c|c|}
\hline
$N_t$ & $N_s$ & $N_0$ & $\beta$ range & temperature range & $\nint$ & $\nconf$ & $\Delta \beta$ \\
\hline
\hline
$6$ &  $96$ &  $48$ & $[5.72785, 5.89985]$ & $[0.7\Tc, \Tc]$ & $3$ & $1.7 \times 10^5$ & $ 10^{-5}$ --- $2\times 10^{-5} $ \\
$6$ &  $96$ &  $48$ & $[5.89985, 6.50667]$ & $[\Tc, 2.5\Tc]$ & $6$ & $3.7 \times 10^5$ & $ 10^{-5}$ --- $2\times 10^{-5} $ \\
\hline
$7$ & $112$ &  $48$ & $[5.79884, 5.98401]$ & $[0.7\Tc, \Tc]$ & $3$ & $2.4 \times 10^5$ & $ 10^{-5} $ \\
$7$ & $112$ &  $48$ &  $[5.98401, 6.6279]$ & $[\Tc, 2.5\Tc]$ & $4$ & $3.3 \times 10^5$ & $ 10^{-5}$ --- $2\times 10^{-5} $ \\
\hline
$8$ & $120$ &  $48$ & $[5.86415, 6.06265]$ & $[0.7\Tc, \Tc]$ & $3$ & $2.6 \times 10^5$ & $ 10^{-5}$ --- $2\times 10^{-5} $ \\
$8$ & $120$ &  $48$ & $[6.06265, 6.72223]$ & $[\Tc, 2.5\Tc]$ & $9$ & $1.2 \times 10^5$ & $ 10^{-5}$ --- $8\times 10^{-5} $ \\
\hline
$10$ & $120$ &  $48$ &  $[5.98408, 6.2068]$ & $[0.7\Tc, \Tc]$ & $5$ & $3.1 \times 10^5$ & $ 7.5\times 10^{-6}$ --- $10^{-5} $ \\
$10$ & $160$ &  $48$ &   $[6.2068, 6.9033]$ & $[\Tc, 2.5\Tc]$ & $8$ & $1.3 \times 10^5$ & $10^{-5}$ --- $10^{-4} $ \\
\hline
\end{tabular}
\caption{Parameters of our simulations.\label{tab:parameters}}
\end{table}

\begin{figure}[h!]
\begin{center}
\includegraphics*[width=\textwidth]{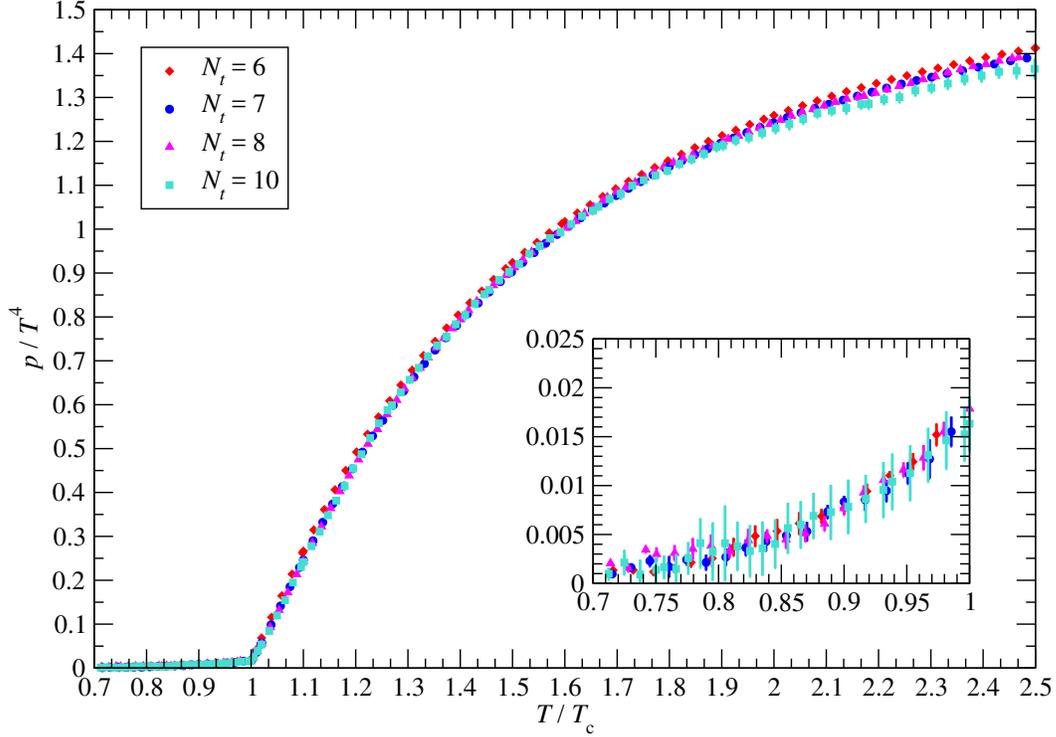}
\caption{\label{fig:finite_Nt_pressure} Results for $p/T^4$, as a function of $T/\Tc$, from simulations at different values of $N_t$. The inset shows a zoom onto the confining phase, $T<\Tc$.}
\end{center}
\end{figure}

The pressure $p$ is the primary thermodynamic observable that we compute using Jarzynski's equality, according to eq.~(\ref{Jarzynski_pressure}): the results at the different values of $N_t$ are shown in fig.~\ref{fig:finite_Nt_pressure}.

\begin{figure}[h!]
\begin{center}
\includegraphics*[width=\textwidth]{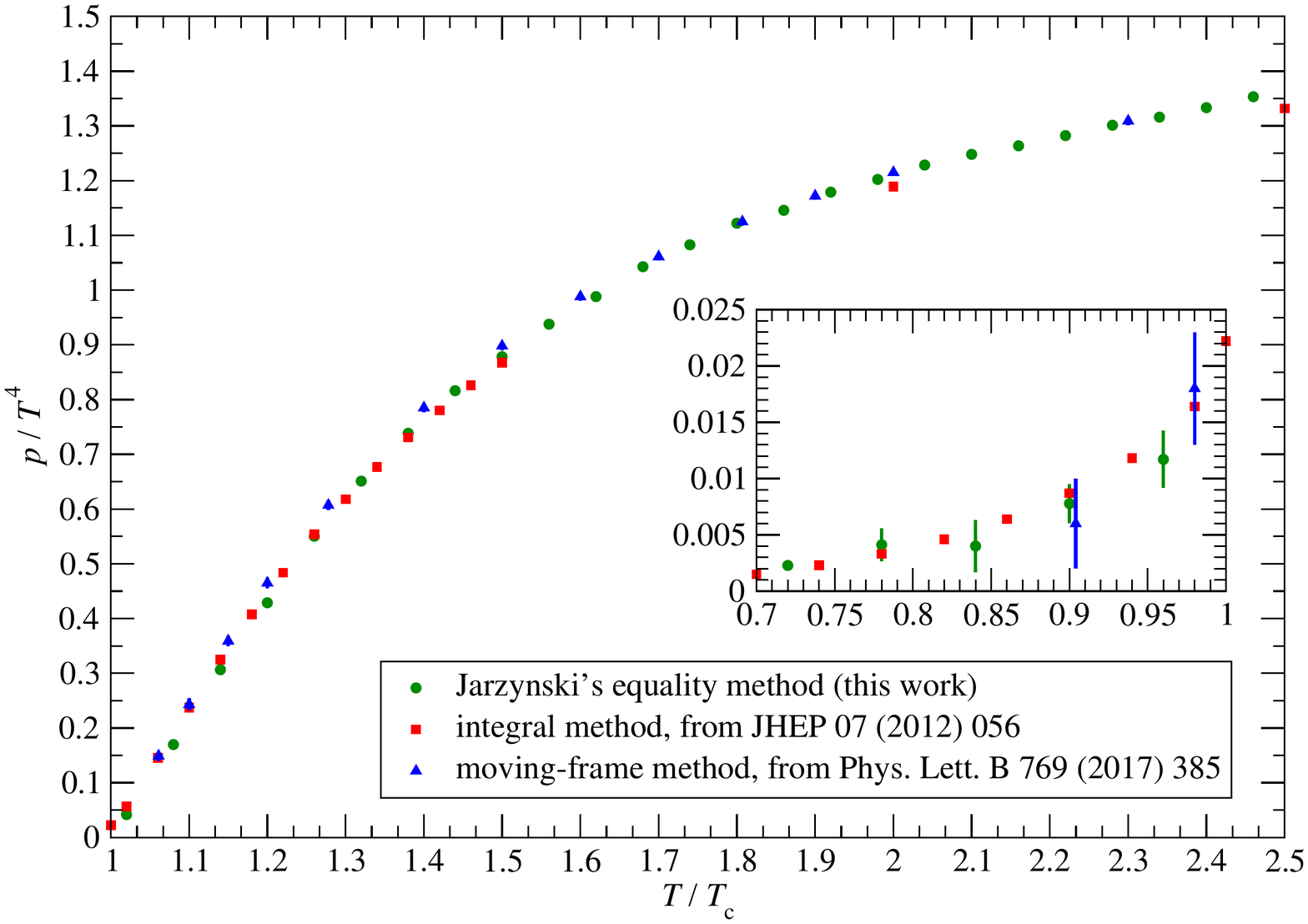}
\caption{\label{fig:pressure} Our results for $p/T^4$, extrapolated to the continuum (green circles), as a function of $T/\Tc$, in comparison with those obtained with the integral method in ref.~\cite{Borsanyi:2012ve} (red squares) and with those obtained using the moving-frame method in ref.~\cite{Giusti:2016iqr} (blue triangles). The results in the confining phase are displayed in the inset plot.}
\end{center}
\end{figure}

From the results for $p/T^4$ at finite lattice spacing, we take the continuum limit by first interpolating them, for each $N_t$, through cubic splines, and then by fitting the splines at fixed values of $T$ with a constant-plus-linear-term fit in $1/N_t^2$:
\begin{equation}
\label{continuum_extrapolation}
p_{N_t}(T) = \alpha(T) + \frac{\xi(T)}{N_t^2}.
\end{equation}
This defines $\alpha(T)$ as the continuum-extrapolated value of the pressure at that temperature. Different types of interpolations at fixed $N_t$, or more complicated functional forms than the one in eq.~(\ref{continuum_extrapolation}), yield compatible results. As the starting value for $p/T^4$ at $\Tin=0.7 \Tc$, we use $p(\Tin)/T_{\mbox{\tiny{in}}}^4 = 0.00086$, the analytical result for a glueball gas~\cite{Hagedorn:1965st} (for a thorough discussion, see also refs.~\cite{Caselle:2015tza, Alba:2016fku} and references therein). Our results for $p/T^4$ obtained in this way are shown in figure~\ref{fig:pressure}, in comparison with those from refs.~\cite{Borsanyi:2012ve, Giusti:2016iqr}.

Other basic thermodynamic observables, like the trace of the energy-momentum tensor $\Delta$, the energy per unit volume $\epsilon$, and the entropy per unit volume $s$ are directly related to the pressure by basic thermodynamic relations:
\begin{eqnarray}
\label{Delta}
\Delta &=& T^5 \frac{\partial}{\partial T} \left( \frac{p}{T^4} \right), \\
\label{energy_density}
\epsilon &=& \frac{T^2}{V} \frac{\partial}{\partial T} \ln Z = 3p + \Delta, \\
\label{entropy_density}
s &=& \frac{\ln Z}{V} + \frac{\epsilon}{T} = \frac{4p + \Delta}{T}.
\end{eqnarray}

\begin{figure}[h!]
\begin{center}
\includegraphics*[width=\textwidth]{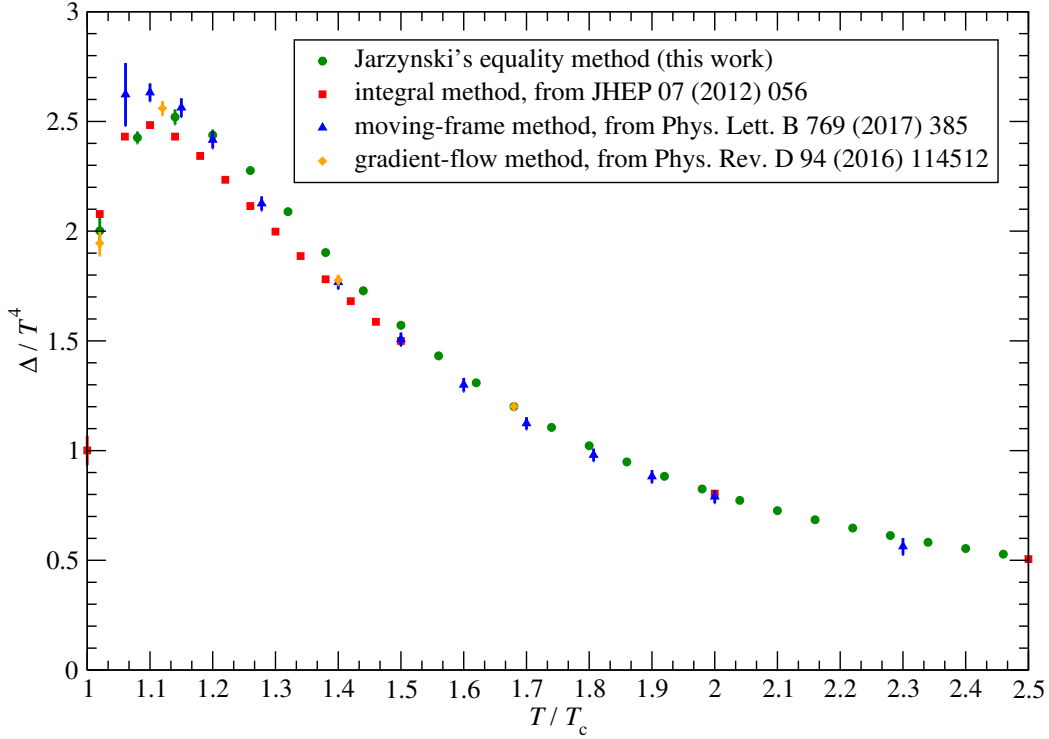}
\caption{\label{fig:Delta} Our continuum-extrapolated results for $\Delta/T^4$ (green circles), as a function of $T/\Tc$, in comparison with those obtained with the integral method in ref.~\cite{Borsanyi:2012ve} (red squares), with those obtained using the moving-frame method in ref.~\cite{Giusti:2016iqr} (blue triangles), and with those computed using the gradient-flow method in ref.~\cite{Kitazawa:2016dsl} (orange diamonds).}
\end{center}
\end{figure}

To compute the trace of the energy-momentum tensor, we first fit our continuum values for $p/T^4$ in the temperature range $\Tc \le T \le 2.5 \Tc$, to the following rational function of $w = \ln(T/\Tc)$:
\begin{equation}
\label{pressure_fit}
\frac{p}{T^4} = \frac{p_1+p_2w+p_3 w^2}{1+p_4w + p_5w^2}.
\end{equation}
The fit gives $p_1=0.0045(35)$, $p_2=1.76(12)$, $p_3=10.6(2.2)$, $p_4=2.07(47)$, and $p_5=5.8(1.1)$, with a reduced $\chi^2$ equal to $0.33$. Deriving the function on the right-hand side of eq.~(\ref{pressure_fit}), we obtain the results for the trace of the energy-momentum tensor shown in fig.~\ref{fig:Delta}, where we compare them with those that have been recently obtained by different groups, using other methods~\cite{Borsanyi:2012ve, Giusti:2016iqr, Kitazawa:2016dsl}.

\begin{figure}[h!]
\begin{center}
\includegraphics*[width=\textwidth]{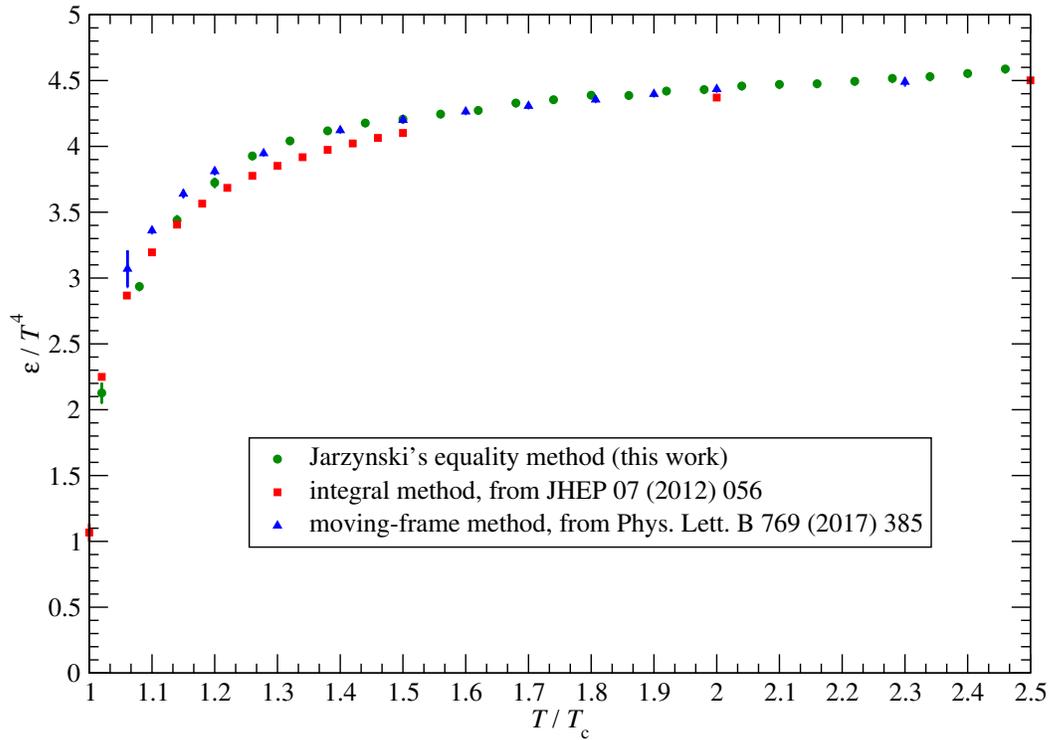}
\caption{\label{fig:energy} Same as in fig.~\protect\ref{fig:pressure}, but for the energy density in units of the fourth power of the temperature.}
\end{center}
\end{figure}

\begin{figure}[h!]
\begin{center}
\includegraphics*[width=\textwidth]{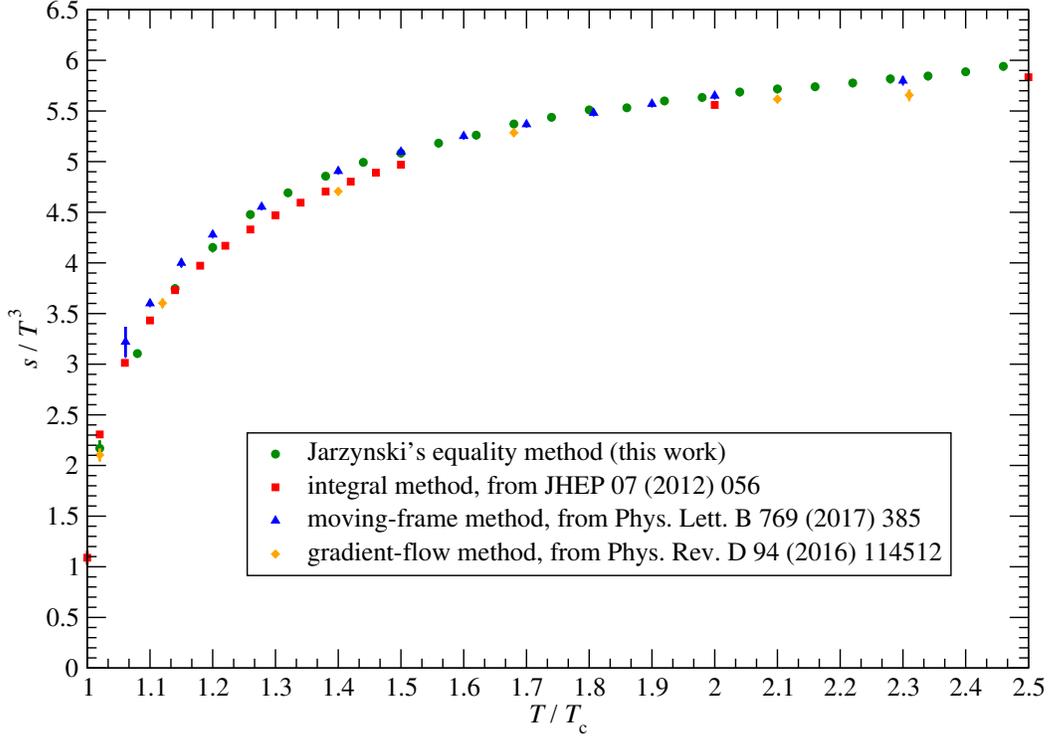}
\caption{\label{fig:entropy} Same as in fig.~\protect\ref{fig:Delta}, but for the entropy density in units of the third power of the temperature.}
\end{center}
\end{figure}

Finally, the energy density and the entropy density are simply obtained using eq.~(\ref{energy_density}) and eq.~(\ref{entropy_density}), respectively: the results are shown in figs.~\ref{fig:energy} and~\ref{fig:entropy}.

The complete set of our continuum-extrapolated results for $p/T^4$, $\Delta/T^4$, $\epsilon/T^4$, and $s/T^3$ is reported in table~\ref{tab:results}.

\begin{table}[!htb]
\centering
\begin{tabular}{|c||c|c|c|c|}
\hline
$T/\Tc$ & $p/T^4$ & $\Delta/T^4$ & $\epsilon/T^4$ & $s/T^3$ \\
\hline
\hline
$0.72$ & $0.0023(3)$  & $0.0137(38)$ & $0.021(17)$ & $0.023(17) $ \\
$0.78$ & $0.0041(15)$ & $0.0200(57)$ & $0.032(20)$ & $0.036(21) $ \\
$0.84$ & $0.0040(23)$ & $0.0286(87)$ & $0.041(23)$ & $0.045(25) $ \\
$0.90$ & $0.0078(17)$ & $0.044(14)$  & $0.067(24)$ & $0.075(26) $ \\
$0.96$ & $0.0117(26)$ & $0.072(23)$  & $0.107(32)$ & $0.119(35) $ \\
$1.02$ & $0.0418(28)$ & $2.001(78)$  & $2.127(78)$ & $2.169(79) $ \\
$1.08$ & $0.1698(32)$ & $2.426(27)$  & $2.936(30)$ & $3.106(32) $ \\
$1.14$ & $0.3064(42)$ & $2.520(34)$  & $3.439(39)$ & $3.746(41) $ \\
$1.20$ & $0.429(6)$   & $2.438(25)$  & $3.724(38)$ & $4.153(43) $ \\
$1.26$ & $0.550(5)$   & $2.276(16)$  & $3.927(27)$ & $4.477(32) $ \\
$1.32$ & $0.651(5)$   & $2.089(14)$  & $4.041(23)$ & $4.693(27) $ \\
$1.38$ & $0.739(5)$   & $1.902(15)$  & $4.118(22)$ & $4.856(26) $ \\
$1.44$ & $0.816(7)$   & $1.728(16)$  & $4.177(24)$ & $4.993(30) $ \\
$1.50$ & $0.878(6)$   & $1.571(16)$  & $4.206(27)$ & $5.084(32) $ \\
$1.56$ & $0.938(7)$   & $1.432(15)$  & $4.245(26)$ & $5.182(31) $ \\
$1.62$ & $0.988(7)$   & $1.309(14)$  & $4.273(26)$ & $5.261(31) $ \\
$1.68$ & $1.043(7)$   & $1.201(12)$  & $4.329(27)$ & $5.372(34) $ \\
$1.74$ & $1.083(6)$   & $1.106(10)$  & $4.355(25)$ & $5.437(30) $ \\
$1.80$ & $1.122(6)$   & $1.022(9)$   & $4.389(25)$ & $5.511(31) $ \\
$1.86$ & $1.146(6)$   & $0.949(8)$   & $4.386(24)$ & $5.532(30) $ \\
$1.92$ & $1.179(7)$   & $0.883(8)$   & $4.420(25)$ & $5.599(32) $ \\
$1.98$ & $1.202(8)$   & $0.825(8)$   & $4.431(27)$ & $5.633(35) $ \\
$2.04$ & $1.229(8)$   & $0.773(8)$   & $4.459(28)$ & $5.687(35) $ \\
$2.10$ & $1.248(8)$   & $0.727(9)$   & $4.471(29)$ & $5.719(37) $ \\
$2.16$ & $1.264(8)$   & $0.685(9)$   & $4.476(27)$ & $5.739(34) $ \\
$2.22$ & $1.282(7)$   & $0.647(10)$  & $4.494(27)$ & $5.776(34) $ \\
$2.28$ & $1.301(8)$   & $0.613(11)$  & $4.516(30)$ & $5.817(37) $ \\
$2.34$ & $1.316(8)$   & $0.582(11)$  & $4.530(29)$ & $5.846(37) $ \\
$2.40$ & $1.333(7)$   & $0.554(12)$  & $4.554(26)$ & $5.887(33) $ \\
$2.46$ & $1.353(7)$   & $0.528(12)$  & $4.588(28)$ & $5.941(35) $ \\
\hline
\end{tabular}
\caption{Our final, continuum-extrapolated results for the pressure (second column), for the trace of the energy-momentum tensor (third column), and for the energy density (fourth column) in units of the fourth power of the temperature, and for the entropy density in units of the third power of the temperature (fifth column), as a function of the temperature in units of the deconfinement temperature (first column).\label{tab:results}}
\end{table}

\section{Discussion and concluding remarks}
\label{sec:discussion}

The results presented in section~\ref{sec:results} deserve several relevant comments, which are separately discussed in each of the following subsections.

\subsection{Universality of lattice results}
\label{subsec:compatibility}

First and foremost, the comparison of our data, obtained with an algorithm based on Jarzynski's equality, with those from previous works~\cite{Borsanyi:2012ve, Giusti:2016iqr, Kitazawa:2016dsl} provides a striking check of the expected universality of lattice results: the fact that the high-precision results obtained by four independent groups, using remarkably different computational strategies, are essentially compatible with each other, indicates that all sources of systematic or statistical uncertainties are under control, and confirms that lattice calculations provide solid, first-principle results for the thermodynamics of strong interactions in the temperature range probed by heavy-ion collision experiments.

Looking at the fine details, however, one can also see that some slight tension between the results obtained with different methods still persists. For example, the results for the various thermodynamic observables reported in ref.~\cite{Borsanyi:2012ve} appear to be systematically lower than the others. This effect is most visible for the trace of the energy momentum tensor in figure~\ref{fig:Delta}, while it is essentially absent in the results for the pressure shown in figure~\ref{fig:pressure} (whereas the energy and entropy densities, being obtained from linear combinations of $p$ and $\Delta$, exhibit an intermediate behavior, with milder tensions). Also, the discrepancy appears to be largest in the temperature region of the peak in $\Delta/T^4$, where it is comparable (in sign and magnitude) with the difference from the results from ref.~\cite{Boyd:1996bx} that was reported in ref.~\cite{Borsanyi:2012ve} itself. While the origin of this slight difference between the results of ref.~\cite{Borsanyi:2012ve} and the others remains unclear,\footnote{It should be noted that ref.~\cite{Borsanyi:2012ve} is the only one, among these works, to use a Symanzik-improved formulation of the lattice action~\cite{Curci:1983an, Luscher:1985zq}, which is affected by smaller discretization effects at finite lattice spacing, and, as a consequence, may offer better control over the extrapolation to the continuum limit. However, we deem it unlikely that the tension with the results reported in the other works can be (completely) interpreted in terms of discretization artifacts, since there is no reason to expect the latter to affect the different quantities, that are computed in those works, in the same way.} it should be remarked that it is quantitatively modest, and does not change the overall physical picture of the $\SU(3)$ equation of state in a significant way.

\subsection{Physical implications of the results}
\label{subsec:physical_implications}

In terms of physics, these results confirm that, in the temperature range relevant for collider experiments, the thermodynamics of $\SU(3)$ Yang-Mills theory is dominated by non-perturbative effects, and far from the ideal limit of a gas of free gluons. In particular, the equilibrium observables considered here are significantly different from their Stefan-Boltzmann values:
\begin{equation}
\label{Stefan-Boltzmann}
\frac{p_{\mbox{\tiny{SB}}}}{T^4} = \frac{8 \pi^2}{45}, \qquad
\frac{\Delta_{\mbox{\tiny{SB}}}}{T^4} = 0, \qquad
\frac{\epsilon_{\mbox{\tiny{SB}}}}{T^4} = \frac{8 \pi^2}{15}, \qquad
\frac{s_{\mbox{\tiny{SB}}}}{T^3} = \frac{32 \pi^2}{45},
\end{equation}
which are reached only in the $T \to \infty$ limit, and approached logarithmically slowly as the temperature is increased. A way to study the values for these quantities at high, but finite, temperatures, is by means of thermal perturbation theory. Weak-coupling expansions for the pressure of QCD (and pure-glue Yang-Mills theory) have a long history: the leading-order correction, $O(g^2)$, was worked out forty years ago~\cite{Shuryak:1977ut, Chin:1978gj}. Soon thereafter, however, it was realized that perturbative expansions in thermal non-Abelian gauge theories have non-trivial features: in particular, the existence of infrared divergences, which have to be resummed, leads to the appearance of terms proportional to \emph{odd} powers and/or logarithms of $g$, and, most importantly, implies that, at some finite order, an \emph{infinite} number of Feynman diagrams, of \emph{arbitrarily complicated topologies}, will contribute~\cite{Linde:1980ts, Gross:1980br}. This ``Linde problem'' leads to the peculiar situation, in which the number of sensible perturbative orders is finite. For the pressure, this problem occurs at $O(g^6)$, or four loops, and the program of computing all perturbative contributions up to that order has been completed, with the determination of all terms $O(g^3)$~\cite{Kapusta:1979fh}, $O(g^4 \ln g)$~\cite{Toimela:1982hv}, $O(g^4)$~\cite{Arnold:1994ps, Arnold:1994eb}, $O(g^5)$~\cite{Zhai:1995ac}, and finally $O(g^6 \ln g)$~\cite{Kajantie:2002wa, Kajantie:2003ax}, but the convergence of the perturbative series is known to be very slow~\cite{Braaten:1995jr, Kajantie:1997tt}. In particular, truncating the perturbative series at subsequent orders results in a strongly oscillating behavior in the temperature range probed in heavy-ion collisions.

As we already mentioned above, dimensional reduction provides an elegant way to systematically account for the non-perturbative physics related to infrared divergences, by means of effective theories~\cite{Ginsparg:1980ef, Appelquist:1981vg} that can be studied non-perturbatively on the lattice~\cite{Kajantie:2000iz, Hietanen:2008tv} (an approach that has recently found useful applications even for real-time phenomena in hot QCD~\cite{CaronHuot:2008ni, Laine:2012ht, Benzke:2012sz, Panero:2013pla, Ghiglieri:2013gia, DOnofrio:2014mld}).

The limited convergence of weak-coupling expansions for thermodynamic quantities in finite-temperature QCD is due to the fact that characteristic phenomena of plasmas, such as screening and Landau damping, must be properly accounted for. To this purpose, one can re-arrange the perturbative expansions using a hard-thermal-loop approach~\cite{Blaizot:2003tw, Andersen:1999fw, Andersen:1999sf, Andersen:2009tc, Andersen:2011sf, Haque:2013sja, Haque:2014rua, Andersen:2015eoa} (in which the Debye mass $m_{\mbox{\tiny{D}}}$ in the ``improvement term'' added to the Lagrangian is, in principle, arbitrary, and must be fixed in a self-consistent way).

In any case, the intrinsically non-perturbative nature of the physics of high-temperature non-Abelian gauge theories makes it hardly surprising that leading-order weak-coupling expansions provide an unsatisfactory description for the equation of state of strong interactions, even at high temperatures. While various phenomenological models (including bottom-up models based on the gauge-gravity duality~\cite{Maldacena:1997re, Gubser:1998bc, Witten:1998qj}) describe well the thermodynamics of the quark-gluon plasma at temperatures close to deconfinement~\cite{Pisarski:2000eq, Meisinger:2001cq, Fukushima:2003fw, Dumitru:2003hp, Ratti:2005jh, Pisarski:2006hz, Vuorinen:2006nz, Kajantie:2006hv, Gursoy:2007cb, Andreev:2007zv, Gursoy:2008bu, Gursoy:2009jd, Alanen:2009xs, Hidaka:2008dr, Fukushima:2008wg, Megias:2009mp, Dumitru:2010mj, Dumitru:2012fw}, lattice calculations remain the most reliable first-principle theoretical tool to study thermal QCD under the conditions probed in heavy-ion collisions.

\subsection{Computational efficiency of the algorithm}
\label{subsec:computational_efficiency}

The algorithmic strategy proposed in ref.~\cite{Caselle:2016wsw} and based on Jarzynski's equality~\cite{Jarzynski:1996oqb, Jarzynski:1997ef} provides a robust and efficient tool to compute the equation of state non-perturbatively on the lattice. As we mentioned above, its implementation in Monte~Carlo calculations only requires that the Markov process satisfies detailed balance, and the assumption that the initial starting configurations (not those at subsequent Monte~Carlo times) are thermalized. It is also interesting to observe that, as we pointed out in section~\ref{sec:results}, in our computations we only used $\ntraj=10$ trajectories for each of the $N_t$ and $N_s$ combinations of values and each of the $\nint$ temperature intervals in the finite-$T$ simulations (and for the corresponding ones at $T=0$). This means that, out of the total number of configurations from which we could extract data for each combination of parameters (which is denoted as $\nconf$ in table~\ref{tab:parameters}, and equals the sum of the $N \cdot \ntraj$ products over the $\nint$ intervals), only a very small number $\nint \cdot \ntraj$ required thermalization.

It is important to discuss the main factors determining the computational efficiency of the algorithm. The key aspect of our algorithm is the exponential average appearing in eq.~(\ref{Jarzynski_theorem}): this implies that, if the typical amplitude of fluctuations in $W/T$ from one trajectory to another is large, then the quantity appearing on the left-hand side of eq.~(\ref{Jarzynski_theorem}) receives its dominant contributions from trajectories in the tail of the distribution of values of $W/T$, and its accurate estimate by Monte~Carlo methods would require a prohibitively large number of trajectories. In the present context,  $W/T$ is replaced by the total variation in Euclidean action $S$ along a trajectory, see eq.~(\ref{action_variation}). Since $S$ (and, as a consequence, $\Delta S$) is an \emph{extensive} quantity, one may expect it to be practically impossible to obtain accurate results on large lattices: in particular, the typical fluctuations in the exponent of eq.~(\ref{Jarzynski_theorem}) will scale like the square root of the lattice hypervolume, making the evaluation of $p(T)$ nearly unattainable on all lattices, except for very coarse ones. Note that this is the same argument by which the sign problem affecting lattice QCD calculations at finite quark chemical potential $\mu$~\cite{deForcrand:2010ys} cannot be solved by the reweighting method~\cite{Ferrenberg:1988yz, Barbour:1997bh, Fodor:2001au}. In fact, the reweighting method is a special case of our algorithm, which reduces to it for $N=1$. Nevertheless, with our algorithm it is possible to take the fluctuations in $\Delta S$ under control even on a lattice of arbitrarily large hypervolume $\mathcal{V}$, for example simply by scaling $N$ proportionally to $\mathcal{V}$. This is so, because the fluctuations in $\Delta S$ from one trajectory to the other result from the sum of the fluctuations in the summands on the right-hand side of eq.~(\ref{action_variation}): assuming that the latter are uncorrelated with each other, when $N$ grows (at fixed $\mathcal{V}$), the fluctuations in $\Delta S$ will be suppressed like $1/\sqrt{N}$ (so that, in particular, in the ``quasi-static limit'' $N \to \infty$ the field configurations remain in equilibrium throughout their evolution from the initial to the final ensemble, and $\Delta S$ is exactly equal to the logarithm of $\Zin/\Zfin$ on all trajectories), thus by making $N$ scale with $\mathcal{V}$, the two, opposite effects on the size of the fluctuations in $\Delta S$ can compensate each other.

A convenient practical way to test whether the numerical results obtained using our algorithm for finite statistics are biased by poor sampling of the distribution of $\Delta S$ values, consists in running the simulation in the direct (from $\lambdain$ to $\lambdafin$) and in the reverse ($\lambdafin \to \lambdain$) direction. If $N$ is too small, then the fluctuations in $\Delta S$ from one trajectory to the other can be very large, and a numerical estimate of the average on the left-hand side of eq.~(\ref{Jarzynski_theorem}) will be determined by a small number of configurations in one of the tails of the distribution, which is very difficult to sample in an accurate way. This will then induce a systematic bias in the numerical results. By carrying out the computation in the reverse direction, the same effect will occur for the variation in Euclidean action induced by a $\lambdafin \to \lambdain$ transformation, but this time for a different distribution, resulting in a generally different bias of numerical results. Thus, an inconsistency in the numerical results obtained from simulations starting from $\lambdain$ or from $\lambdafin$ provides a useful detector of poor-sampling effects.

Note that the mutual consistency of results obtained from a calculation in the direct and in the reverse direction is not a sufficient but a necessary condition for the correctness of the result. All results in our present work pass this test.

More in general, a systematized study of statistical and systematic uncertainties, with the goal of algorithm optimization, can rest on the mathematical theory that has been developed over several years, for generic Monte~Carlo calculations using Jarzynski's equality in statistical mechanics. The ``good practices'' underlying simulations with non-equilibrium work methods are by now well-established, and are encoded in formulas related to the deep connections between statistical mechanics and information theory~\cite{Kullback:1951oi, Cover:2006ei}. For a detailed discussion of the computational efficiency of algorithms based on Jarzynski's equality, see refs.~\cite{Jarzynski:2006re, Pohorille:2010gp, YungerHalpern:2016no, YungerHalpern:2016hm}.

It is interesting to compare the numerical efficiency of our algorithm with other computational methods, that have been used in the literature to evaluate the QCD equation of state on the lattice. Among the three recent works that we directly compared our results with~\cite{Kitazawa:2016dsl, Giusti:2016iqr, Borsanyi:2012ve}, the one reported in ref.~\cite{Borsanyi:2012ve} is based on the most similar method, i.e. the integral method~\cite{Engels:1990vr}. Like for the integral method, our determination of the equation of state is based on the $p=-f$ identity, and requires the numerical subtraction of contributions from quantum, non-thermal fluctuations that would make the free-energy density divergent in the continuum limit $a \to 0$. Exactly like for the integral method, this ultraviolet divergence can be removed by subtracting the free-energy density evaluated at $T=0$ and \emph{at the same lattice spacing}: see the subtrahend in the brackets on the right-hand side of eq.~(\ref{Jarzynski_pressure}). Thus, our method, in itself, does not allow one to bypass the need of renormalization arising in the integral method~\cite{Engels:1990vr}, as it relies on the same vacuum-contribution subtraction. However, an important difference between our method and the integral method is that, while in the latter all field configurations produced at intermediate temperatures (or, equivalently, at intermediate values of $a$, that means at intermediate values of $\beta$) must be fully thermalized, this is not the case for the algorithm based on Jarzynski's equality, in which only the configurations at the initial $\beta$ are thermalized, while those at intermediate (and at the final) $\beta$ values are genuinely out of equilibrium. This implies a significant reduction in CPU time for the algorithm based on Jarzynski's equality.

More quantitatively, as we mentioned above, the thermalization that in this work was used for the configurations at the initial $\beta$ values consisted of $\ntherm = 1.5 \times 10^4$ sweeps (where, as we mentioned above, by ``sweep'' we mean the combination of one heat-bath~\cite{Creutz:1980zw, Kennedy:1985nu} and five to ten over-relaxation updates~\cite{Adler:1981sn, Brown:1987rra} on all link variables of the lattice) for the lattices at finite temperature, and of $\ntherm = 5 \times 10^3$ sweeps for those at $T=0$. Na\"{\i}vely, if one were to make a comparison with a computation of the equation of state based on the integral method~\cite{Engels:1990vr} using the same number of configurations for each data set (the parameter $\nconf$ reported in table~\ref{tab:parameters}), the fact that in our calculation the intermediate configurations need not to be thermalized, would imply a very large reduction in CPU time, by a factor of the order of $N$, the number of steps one trajectory consists of. This estimate of the computational-cost reduction, however, neglects the inherently different nature of the field configurations that are used by the two algorithms. The point is that, while the integral algorithm only uses thermalized configurations, and extracts information on the thermodynamic equilibrium ensemble they belong to, a computation based on Jarzynski's equality attempts to extract information from configurations that \emph{are not} typical ones of the ``target'' equilibrium ensemble (the one specified by the partition function $\Zfin$): in fact, most of them are not typical configuration of \emph{any} equilibrium ensemble, since, by definition of the algorithm, they are not required to thermalize. In practice, the algorithm ``tries to sample'' the target equilibrium ensemble by progressively driving the thermalized configurations of the initial ensemble towards the target ensemble. As remarked above, if the fluctuations in $\Delta S$ are too large, then such sampling becomes computationally very demanding (like in the reweighting method) and exponentially increasing statistics is required for a given level of precision: this is a general feature of all Monte~Carlo algorithms based on Jarzynski's equality, which was shown and discussed in mathematical detail in refs.~\cite{Jarzynski:2006re, Pohorille:2010gp} and, more recently, in refs.~\cite{YungerHalpern:2016no, YungerHalpern:2016hm}, and we refer the interested readers to those references. Note that large fluctuations in $\Delta S$ may occur when $N$ is small, when $\betafin-\betain$ is large, when for $\beta=\betain$ and for $\beta=\betafin$ the system is in two different phases,\footnote{Note that, precisely because of this reason, in our calculations we never used trajectories that crossed the deconfinement phase transition at $T=\Tc$: see table~\ref{tab:parameters}.} or when the number of degrees of freedom is large (including, in particular, when the volume is large); the fact that the fluctuations in $\Delta S$ become large when $\betafin-\betain$ is large (or, more precisely, when the equilibrium statistical ensembles respectively corresponding to $\beta=\betain$ and to $\beta=\betafin$ have little overlap) implies that a proper sampling of such ``long'' trajectories requires higher statistics. Conversely, if the number of steps in each trajectory is increased to large values (with the initial and final parameters fixed), then the simulation proceeds through a sequence of steps which are ``only slightly'' out of equilibrium, and for infinite $N$ the simulation goes through a sequence of configurations in thermal equilibrium.

\begin{figure}[h!]
\begin{center}
\includegraphics*[width=\textwidth]{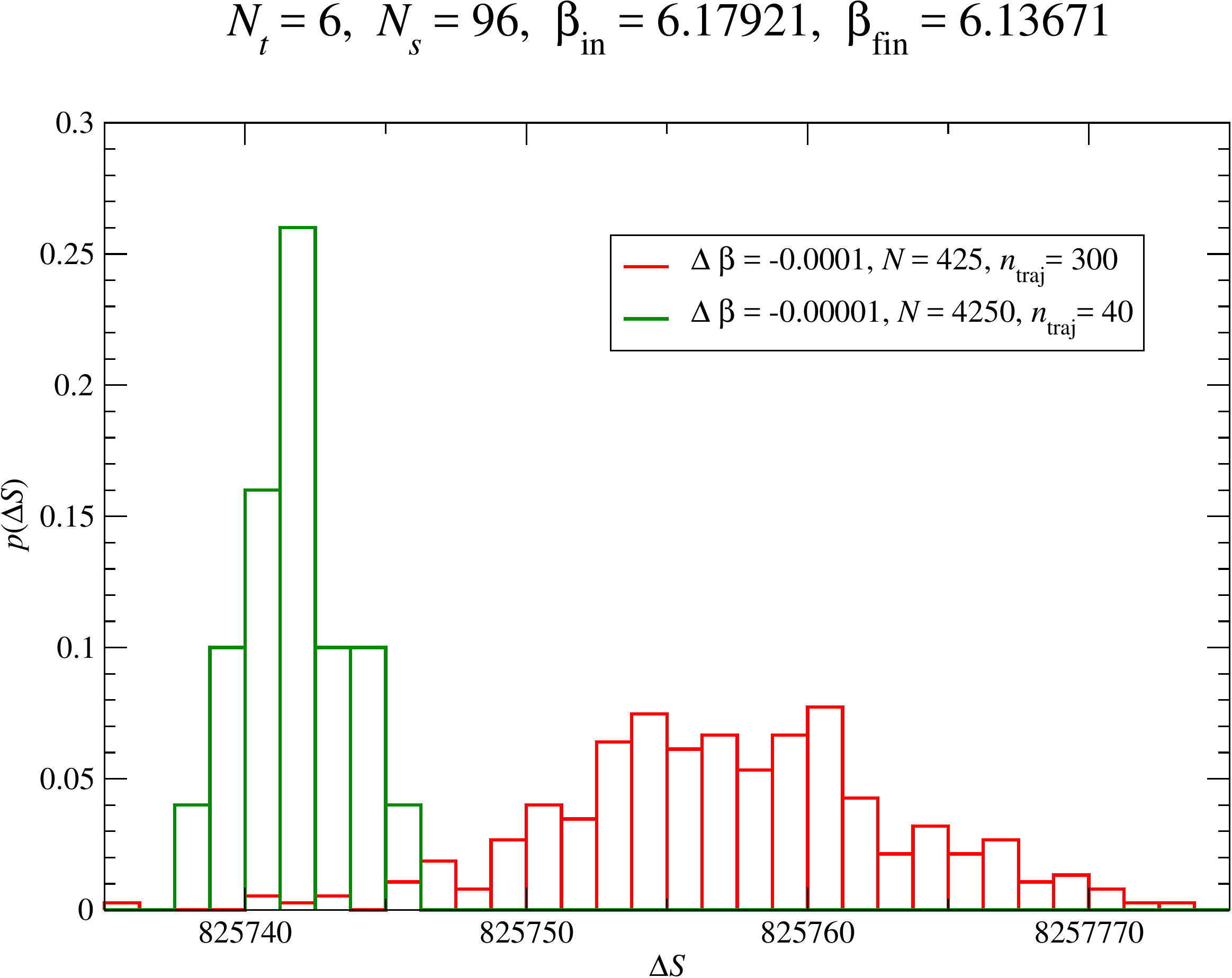}
\caption{\label{fig:histograms_Delta_S_Ncol_3_nt_6_nx_96_ny_96_nz_96_beta_6.1792100000} Histograms of the probability density $p(\Delta S)$ for a variation $\Delta S$ in Euclidean action during non-equilibrium trajectories, as defined in eq.~(\ref{action_variation}), in two different calculations at finite temperature, on a lattice with $N_t=6$ and $N_s=96$. In both simulations, the starting configuration in each of the trajectories is drawn from an equilibrated distribution at $\beta=\betain=6.17921$, and is let evolve through a sequence of non-equilibrium transformations, during which the $\beta$ parameter is decreased down to $\betafin=6.13671$. In the first simulation (red histogram), the non-equilibrium transformations were carried out by dividing the $(\betafin-\betain)$ interval into $N=425$ sub-intervals of amplitude $\left|\Delta \beta\right|=\left|-10^{-4}\right|$ and $300$ trajectories were produced, whereas in the second simulation (green histogram), the $(\betafin-\betain)$ interval was divided into $N=4250$ sub-intervals of amplitude $\left|\Delta \beta\right|=\left|-10^{-5}\right|$, and the number of trajectories was $\ntraj=40$. The estimates for the free-energy difference (in units of the temperature) obtained using the algorithm based on Jarzynski's equality from these two simulations are $\Delta F/T = 825741.5 \pm 4.1$ for the calculation with $N=425$, and $\Delta F/T = 825740.3 \pm 0.5$ for the one with $N=4250$.}
\end{center}
\end{figure}

In order to further clarify the meaning of the non-equilibrium transformations used in simulations based on Jarzynski's theorem, it is instructive to look at examples of the distributions for $\Delta S$, the total Euclidean-action variation during a non-equilibrium trajectory, defined by eq.~(\ref{action_variation}), that can be obtained in simulations starting from the same initial ensemble (at equilibrium), aimed at the same target ensemble, and with a similar computational cost, but with different values of $N$. To this purpose, in fig.~\ref{fig:histograms_Delta_S_Ncol_3_nt_6_nx_96_ny_96_nz_96_beta_6.1792100000} we show the 
density of probability $p$ of observing a variation of Euclidean action $\Delta S$, defined by eq.~(\ref{action_variation}), as obtained from two different simulations on a finite-$T$ lattice with $N_t=6$ and $N_s=96$. More precisely, the histograms display the probability distribution in terms of ``left-stairs'' columns, associated with bins of width $1.25$, whose total area is normalized to one. 

For both calculations, $\betain=6.17921$ and $\betafin=6.13671$, and also the number of configurations used and the total CPU time that was needed to produce them are comparable, but for one of them (whose results are denoted by red histograms) the $(\betafin-\betain)$ interval was split into $425$ intervals, with $\Delta \beta = -10^{-4}$, while for the other one (represented by the green histograms) $\Delta \beta$ was ten times smaller, and $N$ was equal to $4250$.

The fact that, in the latter case, $\Delta \beta$ is much closer to zero implies that the simulation proceeds through a sequence of configurations which are driven out of equilibrium very slowly. As a consequence, one expects the observed distribution of $\Delta S$ values to be close to a very narrow, Gau{\ss}ian-like distribution centered around the free-energy difference (in units of $T$) between the two \emph{equilibrium} ensembles corresponding to $\beta=6.17921$ and $\beta=6.13671$, that one could compute by standard Monte~Carlo calculations at equilibrium on this finite lattice. Indeed, the green histogram plotted in fig.~\ref{fig:histograms_Delta_S_Ncol_3_nt_6_nx_96_ny_96_nz_96_beta_6.1792100000} does confirm this expectation: the distribution is sharply peaked around a value of $\Delta S$ close to $825740$---a value which, unsurprisingly, is fully compatible with the value of $\Delta F/T$ extracted from this simulation using our algorithm based on Jarzynski's theorem: $\Delta F/T = 825740.3 \pm 0.5$. In fact, it is trivial to observe that, when the probability density of $\Delta S$ values tends to a very sharply peaked, nearly $\delta$-like, distribution, then the value of $\Delta F/T$ obtained from eq.~(\ref{Jarzynski_theorem}) (in which, as we stated above, $\Delta S$ plays the r\^ole of $W/T$) coincides with the value of $\Delta S$ at which the peak is located.\footnote{Note that the values of $F/T$ discussed here are \emph{not} renormalized (i.e. the vacuum contribution has not been subtracted yet).}

Much less trivial, however, is the fact that \emph{exactly the same result} is obtained (within statistical uncertainties) when $\Delta F/T$ is calculated by Jarzynski's theorem through the former sample of trajectories, i.e. those obtained with $N=425$ and a significantly larger $\Delta \beta = -10^{-4}$. In this case, $\beta$ is let interpolate from $\betain$ to $\betafin$ at a rate that is ten times faster than in the previous case: as a consequence, the configurations generated by the Monte~Carlo along each trajectory are driven out of equilibrium much more briskly, and, in general, the $\Delta S$ values computed in each non-equilibrium trajectory will fluctuate more wildly (and, in general, in a non-universal, and not trivially predictable, way). Once again, this is clearly visible in our data: the red histogram shows that in this case the distribution of $\Delta S$ values is quite broad, and appears to have a non-trivial structure (even featuring secondary peaks, etc.). From the plot, one also notes that this distribution takes its largest values in the (approximate, and poorly defined) range of $\Delta S$ between $825753$ and $825761$. Remarkably, however, the result for $\Delta F/T$ obtained using Jarzynski's theorem with this set of trajectories is $\Delta F/T = 825741.5 \pm 4.1$, which is very far from the interval where this $p(\Delta S)$ is largest, and perfectly compatible with the one obtained from the set of trajectories with $N=4250$, that are much closer to equilibrium!

It is also worth noting that this result for $\Delta F/T$ has very high precision, of the order of a few per million, comparable with the one achieved in simulations near equilibrium, even though it arises from the exponential average of a quantity (the action variation during non-equilibrium trajectories) whose distribution is so broad. Once again, we remark that, while the details of such distribution may be affected by non-universal dynamics of the Monte~Carlo, with a sizable impact on results obtained from limited statistics, the determination of $\Delta F/T$ through Jarzynski's theorem becomes \emph{exact} when the algorithm samples $p(\Delta S)$ to a sufficient level of precision: the equality encoded in eq.~(\ref{Jarzynski_theorem}) allows one to extract equilibrium information from ensembles of configurations out of equilibrium.

\begin{figure}[h!]
\begin{center}
\includegraphics*[width=\textwidth]{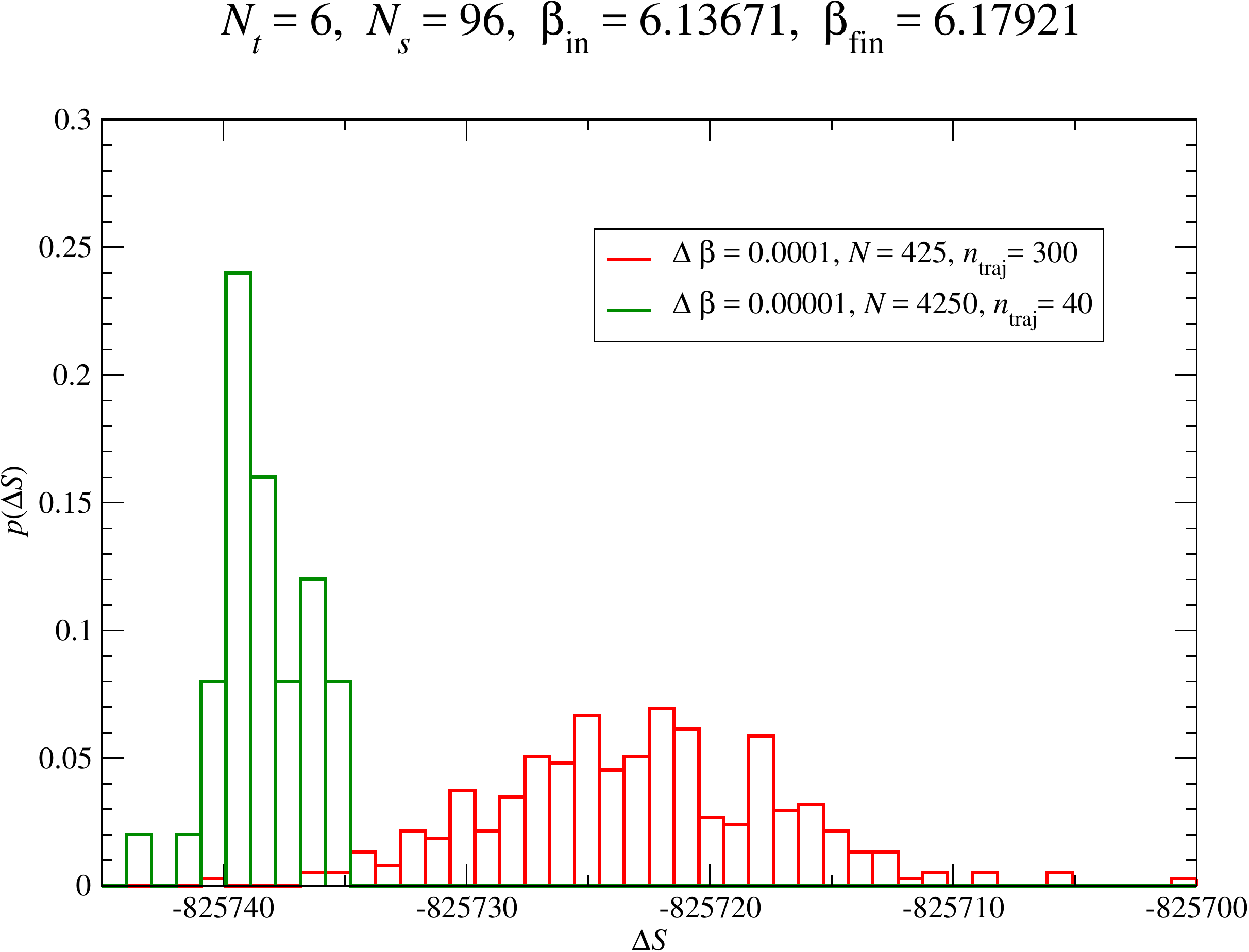}
\caption{\label{fig:histograms_Delta_S_Ncol_3_nt_6_nx_96_ny_96_nz_96_beta_6.1367100000} Same as in fig.~\ref{fig:histograms_Delta_S_Ncol_3_nt_6_nx_96_ny_96_nz_96_beta_6.1792100000}, but for transformations in which $\beta$ is let vary from $\betain=6.13671$ to $\betafin=6.17921$. The red histogram shows the distribution of $\Delta S$ values obtained in simulations with $N=425$ and $\Delta \beta=10^{-4}$ (with $300$ trajectories), whereas the green histogram displays the probability density for the $\Delta S$ values obtained in simulations with $N=4250$ and $\Delta \beta=10^{-5}$, for which $40$ trajectories were produced. The results for the free-energy difference (in units of the temperature) from these two simulations are $\Delta F/T = -825734.7 \pm 2.6$, for the calculation with $N=425$, and $\Delta F/T = -825740.3 \pm 1.1$, for the one with $N=4250$.}
\end{center}
\end{figure}

For completeness, in fig.~\ref{fig:histograms_Delta_S_Ncol_3_nt_6_nx_96_ny_96_nz_96_beta_6.1367100000} we also plot the results obtained from two analogous simulations, carried out \emph{in the opposite direction}, i.e. starting from thermalized configurations at $\beta=6.13671$, and progressively driving  the system out of equilibrium, through a sequence of configuration updates in which $\beta$ is increased to $\beta=6.17921$: like in the previous case, the distribution of $\Delta S$ values obtained at smaller $N$ is the broader and farther from equilibrium one, but the final results for $\Delta F/T$, which in this case are $\Delta F/T = -825734.7 \pm 2.6$ for the simulation with $N=425$, and $\Delta F/T = -825740.3 \pm 1.1$ for the one with $N=4250$, are compatible with each other, and with (minus) the results for $\Delta F/T$ obtained in the simulations with $\betain=6.17921$ and $\betafin=6.13671$, discussed above.

One may also compare our algorithm to compute the pressure, with a variant of the standard integral method combined with the ``snake algorithm'' defined in ref.~\cite{deForcrand:2000fi}, whereby a ratio of partition functions of the form $Z_{\lambdafin}/Z_{\lambdain}$ is factorized into a telescoping product of the form
\begin{equation}
\label{snake_algorithm}
\frac{Z_{\lambdafin}}{Z_{\lambdain}} = \prod_{n=0}^{N-1} \frac{Z_{n+1}}{Z_n}
\end{equation}
(with $Z_0=Z_{\lambdain}$ and $Z_N=Z_{\lambdafin}$), where each $(Z_n,Z_{n+1})$ pair appearing in the intermediate ratios describes statistical ensembles with better overlap than $Z_{\lambdafin}/Z_{\lambdain}$.\footnote{We are indebted to Michele~Pepe for suggesting this comparison to us, and for helpful discussions on the subject.} Note that, for later convenience, we assumed the product on the right-hand side of eq.~(\ref{snake_algorithm}) to involve exactly the same number of terms as the sum on the right-hand side of eq.~(\ref{action_variation}), i.e. $N$. The fundamental idea underlying the snake algorithm is that, when the distributions of configurations associated with $Z_{\lambdain}$ and $Z_{\lambdafin}$ are poorly overlapping, so that a Monte~Carlo estimate of the $Z_{\lambdafin}/Z_{\lambdain}$ ratio to a fixed level of relative precision would require exponentially large statistics, evaluating each of the $Z_{n+1}/Z_n$ ratios appearing on the right-hand side of eq.~(\ref{snake_algorithm}) is computationally much cheaper, provided $Z_n$ and $Z_{n+1}$ always describe ensembles with a good overlap with each other. Then, all $Z_{n+1}/Z_n$ ratios are $O(1)$, and can be evaluated to high precision with a fixed computational cost. The final statistical uncertainty on $Z_{\lambdafin}/Z_{\lambdain}$ is eventually obtained by the sum (in quadrature, as they are obtained from independent simulations) of the uncertainties on the $Z_{n+1}/Z_n$ ratios, and does not grow exponentially. If $Z_{\lambdafin}/Z_{\lambdain}$ is factorized in exactly $N$ terms (as we assumed in the equation above), and if each of the $Z_{n+1}/Z_n$ ratios is calculated by Monte~Carlo methods with $\ntraj$ configurations, then with the snake algorithm one would be able to determine $Z_{\lambdafin}/Z_{\lambdain}$ by producing $N \cdot \ntraj$ \emph{thermalized} and \emph{uncorrelated} configurations. An elementary argument shows that, while with this number of independent configurations the algorithm based on Jarzynski's equality would yield an estimate of $Z_{\lambdain}/Z_{\lambdafin}$ from $\ntraj$ measurements, the snake algorithm would instead express the same quantity as
\begin{equation}
\label{snake_explicit}
\prod_{n=0}^{N-1} \left[ \frac{1}{\ntraj} \sum_{k_n=0}^{\ntraj} \left( \frac{Z_{n+1}}{Z_n}\right)_{k_n} \right]
\end{equation}
(where $(Z_{n+1}/Z_n)_{k_n}$ denotes the value of the $Z_{n+1}/Z_n$ ratio computed in the $k_n$-th configuration), which, when the product is expanded, corresponds to ${\ntraj}^N$ measurements. While this na\"{\i}ve counting argument overlooks the r\^ole of fluctuations, it suggests that, under these conditions (i.e. working with a sample of $N \cdot \ntraj$, completely thermalized and fully independent configurations), the snake algorithm would outperform the one based on Jarzynski's equality. This is not surprising: indeed, were the Markov update algorithm perfectly efficient, i.e. capable of generating fully thermalized and decorrelated configurations in a single sweep, then the system would never be out of equilibrium. In that case, our algorithm would not be affected by any overlap problem even if the $ \lambdain \to \lambdafin $ transformation were carried out in a single step: this means that our algorithm could be reduced to the reweighting algorithm, as discussed above, and, with a sample of $N \cdot \ntraj$ configurations, one could factor the $\Zfin/\Zin$ ratio into a product of $N$ intermediate ratios, each of which could be computed by reweighting. Clearly, however, it is in the more realistic case of Markov updates that do not produce immediate thermalization, that our simulation algorithm reveals its full potential: in this case, our algorithm turns the fact that the field configurations ``lag behind'' equilibrium into an advantage, by means of Jarzynski's equality---whereas a computation based on the standard integral method, or on its snake-algorithm variant, would always require thermalized configurations, which would increase its computational cost. 

As concerns the two other works that we confronted our results with~\cite{Kitazawa:2016dsl, Giusti:2016iqr}, a comparison of the computational costs is far less direct.

In the calculation of the equation of state presented in ref.~\cite{Kitazawa:2016dsl}, based on the Wilson flow~\cite{Luscher:2010iy}, the energy density and the pressure are extracted from the diagonal components of the renormalized energy-momentum tensor of the theory, which, in turn, is obtained from the behavior of the two dimension-four, gauge-invariant operators defined in terms of the flowed field-strength tensor at short flow time~\cite{Suzuki:2013gza}. The calculation involves the numerical solution of the differential equation defining the flowed gauge field, and a double extrapolation, in which the small-flow-time limit has to be taken after the continuum-limit ($a \to 0$) extrapolation.

Finally, in ref.~\cite{Giusti:2016iqr}, the primary observable to determine the equation of state is the entropy density (in units of $T^3$), which is directly related to the time-space off-diagonal components of the energy-momentum tensor. When the theory is defined in a relativistic moving frame, it is possible to prove a set of Ward-Takahashi identities for the correlators of the energy-momentum tensor, that relate the energy and momentum distributions in the canonical ensemble, and allow one to non-perturbatively renormalize the energy-momentum tensor~\cite{Giusti:2012yj}. In practice, this multiplicative renormalization of the energy-momentum tensor is encoded in a finite function of the bare coupling, which has to be determined independently.

\subsection{Further applications of the algorithm based on Jarzynski's equality}
\label{subsec:extensions}

Extending our algorithm to calculations including dynamical quark flavors is straightforward, and we plan to implement it in code for lattice simulations of full QCD in future work. In this respect, it would be interesting to compare the efficiency of this algorithm to different calculations of the QCD equation of state~\cite{Borsanyi:2013bia, Bazavov:2014pvz, Aoki:2009sc, Bazavov:2011nk, Bhattacharya:2014ara, Burger:2014xga, Taniguchi:2016ofw, Kanaya:2016rkt, DallaBrida:2017sxr}.

Another direction, in which the present work can be generalized, consists in applying the Jarzynski's equality to lattice calculations of different physical observables. The computational strategy based on this algorithm, indeed, is quite general and versatile, and not restricted to the thermodynamics domain. As a benchmark study, a determination of the interface free energy was presented in ref.~\cite{Caselle:2016wsw}; the extension to other quantities, like the running coupling of the strong interaction in the Schr\"odinger-functional scheme~\cite{Luscher:1991wu, Luscher:1992an, Luscher:1992zx, Luscher:1993gh, Sint:1993un, Sint:1995ch, Bode:1998hd, Bode:1999sm} and the entanglement entropy in lattice gauge theory~\cite{Buividovich:2008gq, Buividovich:2008kq, Velytsky:2008rs, Donnelly:2011hn, Aoki:2015bsa, Radicevic:2015sza, Itou:2015cyu}, is under way.

\vskip1.0cm 
\noindent{\bf Acknowledgements}\\
The simulations were run on the supercomputers of the Consorzio Interuniversitario per il Calcolo Automatico dell'Italia Nord Orientale (CINECA). We thank Mattia~Dalla~Brida, Leonardo~Giusti, Martin~Hasenbusch, Michele~Pepe, Antonio~Rago, and Rainer~Sommer for helpful comments and discussions.

\bibliography{paper}

\end{document}